\documentclass[reprint, superscriptaddress, floatfix, amsmath,amssymb, aps, ]{revtex4-2}

\usepackage[utf8]{inputenc}
\usepackage{amsmath,esint}
\usepackage{mathtools}
\usepackage[colorlinks=true,linkcolor=blue,citecolor=blue,urlcolor=blue]{hyperref}
\usepackage{amsfonts}
\usepackage{amssymb}
\usepackage{bm}
\usepackage{textcomp}
\usepackage{empheq}
\usepackage{siunitx}
\usepackage[titletoc,title]{appendix}
\usepackage{graphicx}
\usepackage{dcolumn}
\usepackage{bm}
\usepackage{xcolor,soul}

\newcommand{\citeasnoun}[1]{Ref.~\cite{#1}}

\newcommand{\figref}[1]{Fig.~\ref{fig:#1}}
\renewcommand{\eqref}[1]{Eq.~(\ref{eq:#1})}
\newcommand{\Eqref}[1]{Equation~(\ref{eq:#1})}

\newcommand{\vect}[1]{\mathbf{#1}}
\newcommand*{\xv}{\vect{x}}
\newcommand*{\Ev}{\vect{E}}
\newcommand*{\ev}{\vect{e}}
\newcommand*{\Hv}{\vect{H}}
\newcommand*{\Einc}{\Ev_{\rm inc}}
\newcommand*{\einc}{\ev_{\rm inc}}
\newcommand*{\Pv}{\vect{P}}
\newcommand*{\pv}{\vect{p}}
\newcommand*{\Jv}{\vect{J}}
\newcommand*{\TT}{\mathbb{T}}
\newcommand*{\II}{\mathbb{I}}

\renewcommand{\Re}{\operatorname{Re}}
\renewcommand{\Im}{\operatorname{Im}}
\newcommand*{\SM}{SM}

\begin{document}

\title{All electromagnetic scattering bodies are matrix-valued oscillators}
\author{Lang Zhang}
\author{Francesco Monticone}
\author{Owen D. Miller}

\date{\today}

\begin{abstract}
    In this article, we introduce a new viewpoint on electromagnetic scattering. Tailoring spectral electromagnetic response underpins important applications ranging from sensing to energy conversion, and is flourishing with new ideas from non-Hermitian physics. There exist excellent theoretical tools for modeling such responses, particularly coupled-mode theories and quasinormal-mode expansions. Yet these approaches offer little insight into the outer limits of what is possible when broadband light interacts with any designable nanophotonic pattern. We show that a special scattering matrix, the ``$\TT$'' matrix, can always be decomposed into a set of fictitious Drude--Lorentz oscillators with matrix-valued (spatially nonlocal) coefficients. For any application and any scatterer, the only designable degrees of freedom are these matrix coefficients, implying strong constraints on lineshapes and response functions that had previously been ``hidden.'' To demonstrate the power of this approach, we apply it to near-field radiative heat transfer, where there has been a long-standing gap between the best known designs and theoretical limits to maximum energy exchange. Our new framework identifies upper bounds that come quite close to the current state-of-the-art, and explains why unconventional plasmonic materials should be superior to conventional plasmonic materials. More generally, this approach can be seamlessly applied to high-interest applications across nanophotonics---including for metasurfaces, imaging, and photovoltaics---and may be generalizable to unique challenges that arise in acoustic and/or quantum scattering theory.
\end{abstract}

\maketitle
Probing and harnessing the frequency dependence of electromagnetic interactions underlies atomic spectroscopy, molecular sensing, information and energy technologies, and more~\cite{Condon1951,Bernath2005,Shannon1948,Shockley1961}. Decades of research into resonant expansions and normal- and quasinormal-mode theories now enable complex scenarios to often be well-described by a small number of ``physical oscillators''~\cite{Ching1998,Tang2000,Fan_tcmt,Sauvan2022}. These physical oscillators offer high-accuracy \emph{descriptive} modeling, but they provide little \emph{prescriptive} guidance: what lineshapes are physically possible, and what are the ultimate limits of corralling broadband radiation?

To address these foundational questions of wave physics, we introduce a new decomposition of the electromagnetic scattering process into an infinite set of fictitious mathematical ``oscillators'' that offer a general framework for prescriptive guidance and fundamental limits. Central to our approach is an arguably under-appreciated scattering operator, the ``$\TT$ matrix''~\cite{Carminati2021}, which relates incident fields to the polarization fields they induce. The complete scattering response of any scattering body is determined by its $\TT$ matrix, but what form can this matrix take? We show that causality and passivity impose strong constraints on its possible form. The culmination of this insight is a powerful and general representation: every scattering body's $\TT$ matrix must be a superposition of lossless Drude--Lorentz oscillators with matrix-valued (spatially nonlocal) coefficients. Moreover, the oscillator coefficients are highly constrained, with three properties absent in other approaches: (1) they are Hermitian, even in lossy and open systems, (2) they have finite sum rules in many scenarios, and (3) they are positive-semidefinite in passive systems. As a whole, we arrive at a striking conclusion: the controllable degrees of freedom of \emph{any} scattering body, for \emph{any} application and spectral range, are \emph{only} the (highly constrained) matrix-valued oscillator strengths. Such limited DOFs must imply strong constraints on scattering response; more generally, this new representation offers a general method for identifying fundamental limits to spectral control.

To demonstrate the power of this approach, we use our oscillator decomposition of $\TT$ matrices to resolve a long-standing question in energy transport: what is the maximum rate at which two bodies can radiatively exchange heat in the near field? Going back many decades, it has been understood that radiative heat exchange in the near field can be substantially larger than its far-field counterpart~\cite{Polder1971,Rytov1988,Volokitin2007}, due to the enormous number of accessible evanescent channels in addition to propagating ones, yet the maximum extent of this enhancement--with ramifications for applications such as thermophotovoltaics~\cite{Fiorino2018,Bhatt2020}, photonic refrigeration~\cite{Zhu2019}, and heat-assisted magnetic recording~\cite{Challener2009}--has been far less clear. Previous theoretical bounds~\cite{Pendry1999,Ben-Abdallah2010,Miller2015,Shim2019bandwidth,limited_role_structuring} have suggested strong material-electron-density dependencies, unbounded response for low-loss materials, and orders-of-magnitude gaps from known designs ($>$ 750X). Using our $\TT$-matrix representation, we offer a new simple, general theoretical limit that is within a small factor (5X) of the state-of-the-art. Encapsulated in our approach is a sum rule that explains why planar structures are better than sharp-tip patterns for large-area, broadband near-field enhancements, and why unconventional plasmonic materials should offer the largest enhancements. 

\section{The refractive-index paradox}
At a microscopic level, accurate computations of the linear optical response (e.g., refractive index) of a material is sufficiently difficult to be at the frontier of modern electronic-structure methods~\cite{Burke2012}. Yet modeling uncertainty does not imply unbounded possibility: there are well-known constraints to allowable refractive index~\cite{Jackson1999,Andreoli2021,Shim_refractive_index}. This understanding comes not from density-functional-theoretic constraints, but instead fundamental principles such as causality and passivity. At the macroscopic level of wave scattering, however, the situation is reversed. With fast-solver techniques pioneered over three decades, it is possible to compute the electromagnetic fields of structures that are thousands of wavelengths in size at machine precision~\cite{Martinsson2019}. Yet our understanding of the \emph{extreme} responses of those same structures is highly limited. Why has our understanding of limits at the macroscopic scale, where computational methods are more successful, not reached our understanding of limits at the microscopic scale? It is because the combined principles of causality and passivity have not been integrated into scattering-matrix representations. This shortcoming can be remedied in the scattering $\TT$ matrix.

\section{The scattering $\TT$ matrix}
In linear, time-invariant electrodynamics, there must be a linear operator that relates electromagnetic fields incident upon a scatterer to the polarization fields they induce; the standard name for this response function is the ``$\TT$ operator.'' For simplicity of notation and exposition, we assume any standard numerical discretization of sufficiently high accuracy such that the $\TT$ operator becomes a $\TT$ matrix; if we collate the incident fields $\Einc(\xv)$ into a vector $\einc$ and the polarization fields $\Pv(\xv)$ into a vector $\pv$, then the frequency-domain relation ($e^{-i\omega t}$ sign convention) defining the corresponding $\TT$ matrix is,
\begin{align}
    \pv(\omega) = \TT(\omega) \einc(\omega),
\end{align}
which is the discrete analog of the convolution equation $\Pv(\xv,\omega) = \int \TT(\xv,\xv',\omega) \Einc(\xv',\omega) \,{\rm d}\xv'$. The $\TT$ matrix can be derived from first principles via integral operators, as discussed in \citeasnoun{Carminati2021} and the {\SM}. The key property of the $\TT$ matrix, for our purposes, is that it is a causal response function: the polarization field at any $\xv$ cannot be excited by an incident field at $\xv'$ until \emph{after} the incident field has reached $\xv'$. This is similar to the causality condition for material susceptibilities~\cite{Nussenzveig1972}, and it has an analogous consequence. Using Fourier-based arguments that parallel those for material susceptibilities, we find a Kramers--Kronig-like equation that must be satisfied by any $\TT$ matrix (detailed proof in SM):
\begin{align}
    \Re \TT(\omega) = \frac{2}{\pi} \int_0^\infty \frac{\omega' \Im \TT(\omega')}{(\omega')^2 - \omega^2} \,{\rm d}\omega'.
    \label{eq:KKTT}
\end{align}
\Eqref{KKTT} is a Kramers--Kronig (or Hilbert transform) relation for arbitrary scattering processes, for any underlying material. An important feature of \eqref{KKTT} is that it is matrix-valued; the ``$\Re$'' and ``$\Im$'' operators denote the Hermitian and anti-Hermitian parts of their arguments, respectively. \Eqref{KKTT} implies that once the anti-Hermitian part of the $\TT$ matrix is known for all frequencies, its Hermitian part has been determined as well, and vice versa. 

Next we identify sum rules for the $\TT$ matrix. There are two special frequencies at which the left-hand side of \eqref{KKTT} simplifies: zero frequency (electrostatics) and infinite frequency (where materials become transparent). Which is more useful depends on the frequency range of a given application; we include here the high-frequency sum rule, and leave the low-frequency sum rule for the {\SM}. At infinite frequency, the electrons of a material can be regarded as free, and material susceptibilities must scale as $\chi(\omega) \rightarrow -\omega_p^2 / \omega^2$, where $\omega_p^2$ is proportional to the total electron density of the material~\cite{Nussenzveig1972}. In this limit, the first Born approximation is asymptotically exact, and the polarization field is given by $\Pv \simeq \chi \Einc \simeq -(\omega_p^2 / \omega^2) \Einc$ (in units where the free-space permittivity is 1), implying that the $\TT$ matrix asymptotically approaches $-(\omega_p^2 / \omega^2) \II$, where $\II$ is the identity matrix. Inserting this limit into the KK relation of \eqref{KKTT} yields the sum rule,
\begin{align}
    \int_0^{\infty} \omega \Im \TT(\omega) \,{\rm d}\omega = \frac{\pi\omega_p^2}{2} \II.
    \label{eq:SRTT}
\end{align}
This sum rule constrains the total contributions from $\Im \TT(\omega)$ over all frequencies, much like the $f$ sum rule for material-susceptibility oscillator strengths~\cite{Reiche1925,Kuhn1925,sum_rule_chi}.

The final piece of the puzzle is passivity. In passive scatterers, the polarization fields do no net work. Mathematically, the work done by the incident fields on the polarization currents $\Jv$ is $\frac{1}{2} \Re \int \Einc^* \cdot \Jv = \frac{\omega}{2} \Im \int \Einc^* \cdot \Pv$; positivity of this expression implies, at positive frequencies, the positive-semidefinite condition, $\Im \TT(\omega) \geq 0$.
\begin{figure*}[htb]
  \includegraphics[width=\textwidth]{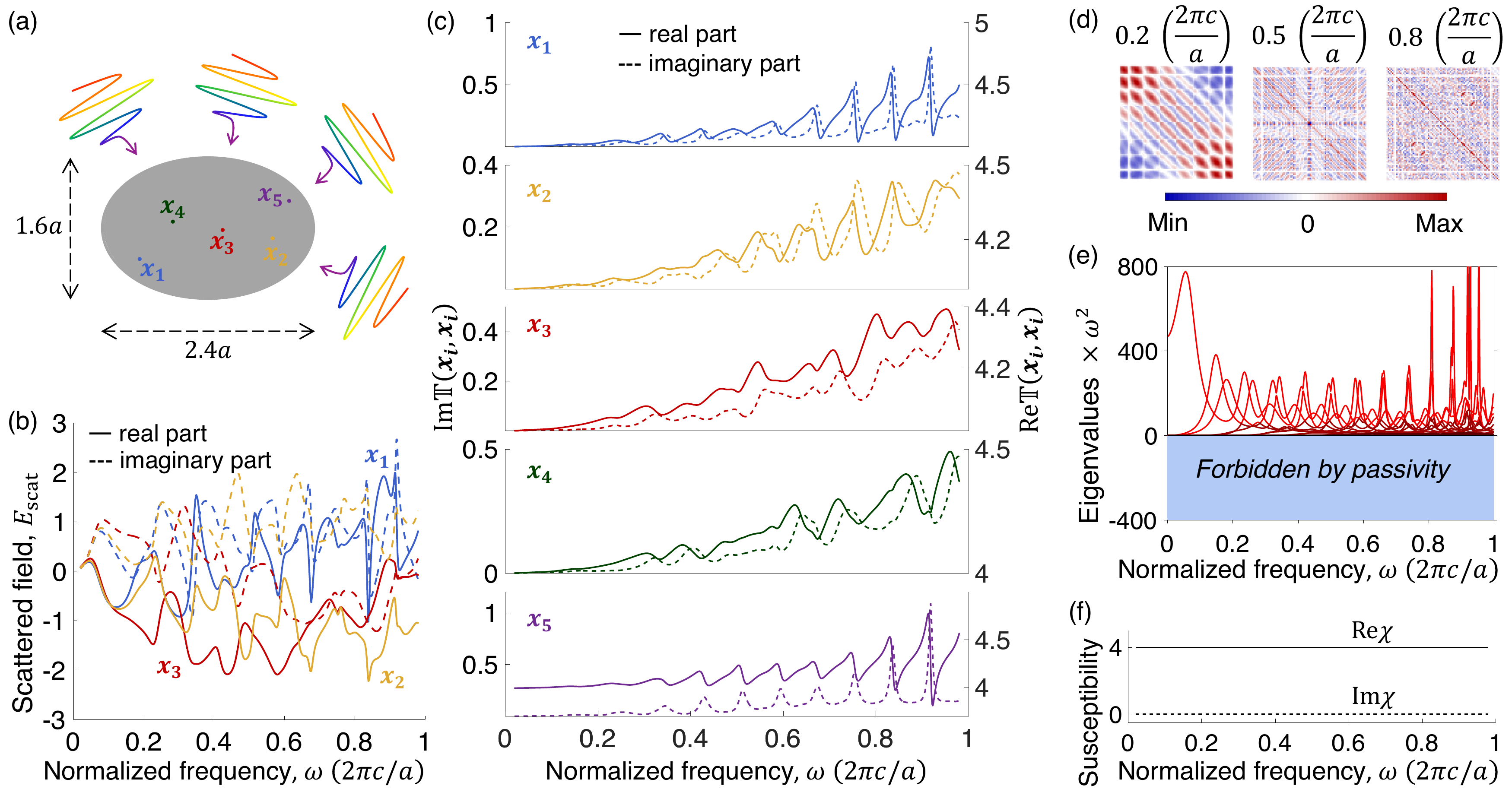}
  \caption{Broadband scattering of incident plane waves from an elliptical cylinder. A schematic depiction is provided in (a). Whereas the scattered fields (at points 1--5 of part a) exhibit seemingly random variations, as depicted in (b) for a single incident angle, the $\TT$ matrix has far more spectral structure. Part (c) shows the real and imaginary parts of the diagonal elements of the $\TT$ matrix, which look reminiscent of the structured oscillations of a material susceptibility. Yet this cylinder has a constant susceptibility $\chi = 4$, as shown in (f). The undulations in the $\TT$ matrix elements are not material oscillators, but a new type of matrix-valued, nonlocal scattering oscillator. Parts (d) and (e) show the matrix-valued scattering-oscillator coefficients $\TT_i$, and their eigenvalues, respectively, showcasing the extent to which passivity and causality restrict the possible form of the scattering $\TT$ matrix.}
  \label{fig:Tscat}
\end{figure*}

\section{Oscillator representation}
We now have three key ingredients: a Kramers-Kronig relation, a sum rule, and a positivity constraint. Together, we can use these to form a new spectral representation of any $\TT$ matrix. Again, for simplicity of exposition, we work in a discrete basis, discussing the straightforward generalization to continuous frequencies in the {\SM}. Consider a discrete set of frequencies $\omega_i$ from 0 to $\infty$ at which we equate $\omega \Im \TT(\omega)$, with appropriate prefactors, to a set of frequency-independent Hermitian matrices $\TT_i$:
\begin{align}
   \omega \Im \TT(\omega) = \frac{\pi\omega_p^2}{2} \sum_i \TT_i \delta(\omega - \omega).
\end{align}
By passivity, all $\TT_i$ are positive semidefinite, $\TT_i \geq 0$, and by the sum rule, constrained to a finite sum: $\sum_i \TT_i \leq \II$. Finally, the KK relation of \eqref{KKTT} dictates that by encoding the anti-Hermitian part of $\TT(\omega)$ into matrix variables $\TT_i$, the Hermitian part of $\TT(\omega)$ is completely determined; there are no more degrees of freedom. Moreover, the spectral lineshape of the Hermitian part around every frequency $\omega_i$ is that of a lossless Drude--Lorentz oscillator. A few mathematical steps can make this connection more rigorous ([SM]), such that the three ingredients are encoded in a single, general representation:
\begin{align}
    \TT(\omega) = \lim_{\gamma\rightarrow 0} \sum_i \frac{\omega_p^2}{\omega_i^2 - \omega^2 - i\gamma\omega} \TT_i, 
    \label{eq:OscTT}
\end{align}
along with the conditions that $\TT_i \geq 0$ for all $i$ and $\sum_i \TT_i = \II$. \Eqref{OscTT} is the foundational result of our paper: the $\TT$ matrix of every linear scattering body must be decomposable into a (fictitious) set of lossless Drude--Lorentz oscillators, with matrix-valued coefficients that are positive-semidefinite and constrained in total strength. This representation is contrasted with standard oscillator decompositions (coupled-mode theory, quasinormal modes) in the SM. In particular, we stress that this is not an eigendecomposition, and $\omega_i$ are not eigenfrequencies of the operator. Given an electron density, the only degrees of freedom in the scattering process are the matrices $\TT_i$. Moreover, the constraints on these matrices ($\TT_i \geq 0$, $\sum_i \TT_i \leq \II$) are convex sets, and the $\TT$ matrix is linear (and thereby convex) in these matrices. Hence, not only does \eqref{OscTT} appear to offer a strong constraint on possible spectral response, but its constituents are ideally suited for mathematical optimization and consequent fundamental limits.

As a first demonstration of the surprising structure implied by this representation, we consider broadband scattering from an elliptical dielectric cylinder. The scattered electric field (total field minus incident field) at various points within the scatterer, computed by full-wave simulations [SM], is shown in \figref{Tscat}(b), but is hard to interpret due to its seemingly random undulations. The tremendous advances in quasinormal-mode (QNM) techniques suggest that one could accurately reproduce these fields with a relatively small number of QNMs~\cite{Sauvan2022}, but that modeling capability does not imply an understanding of the extreme limits of what is possible. How many resonances can be excited? With what amplitudes, phases, and overlaps with power-carrying channels? 

By contrast, consider the diagonal elements of the real (Hermitian) and imaginary (anti-Hermitian) parts of the $\TT$ matrix, as depicted in \figref{Tscat}(c). They show the characteristic features of oscillators! The lineshapes of the $\TT$-matrix elements mimic exactly the Drude--Lorentz-like behavior of electronic transitions, but they arise not from real material oscillators (the susceptibility is constant, \figref{Tscat}(f)), but from complex wave-scattering behavior itself. Included in \figref{Tscat}(d) are the matrix-valued coefficients at three frequencies, with the eigenvalues of all frequencies plotted in \figref{Tscat}(e). They are all positive, as guaranteed by passivity. By uncovering this structure, we can now see that scattering processes must be decomposable into a collection of positive-semidefinite matrices with finite total strength. Such decompositions are ideally suited for identifying fundamental limits to spectral control. 

\section{Ultimate limits to NFRHT}
\begin{figure*}[htb]
  \includegraphics[width=\textwidth]{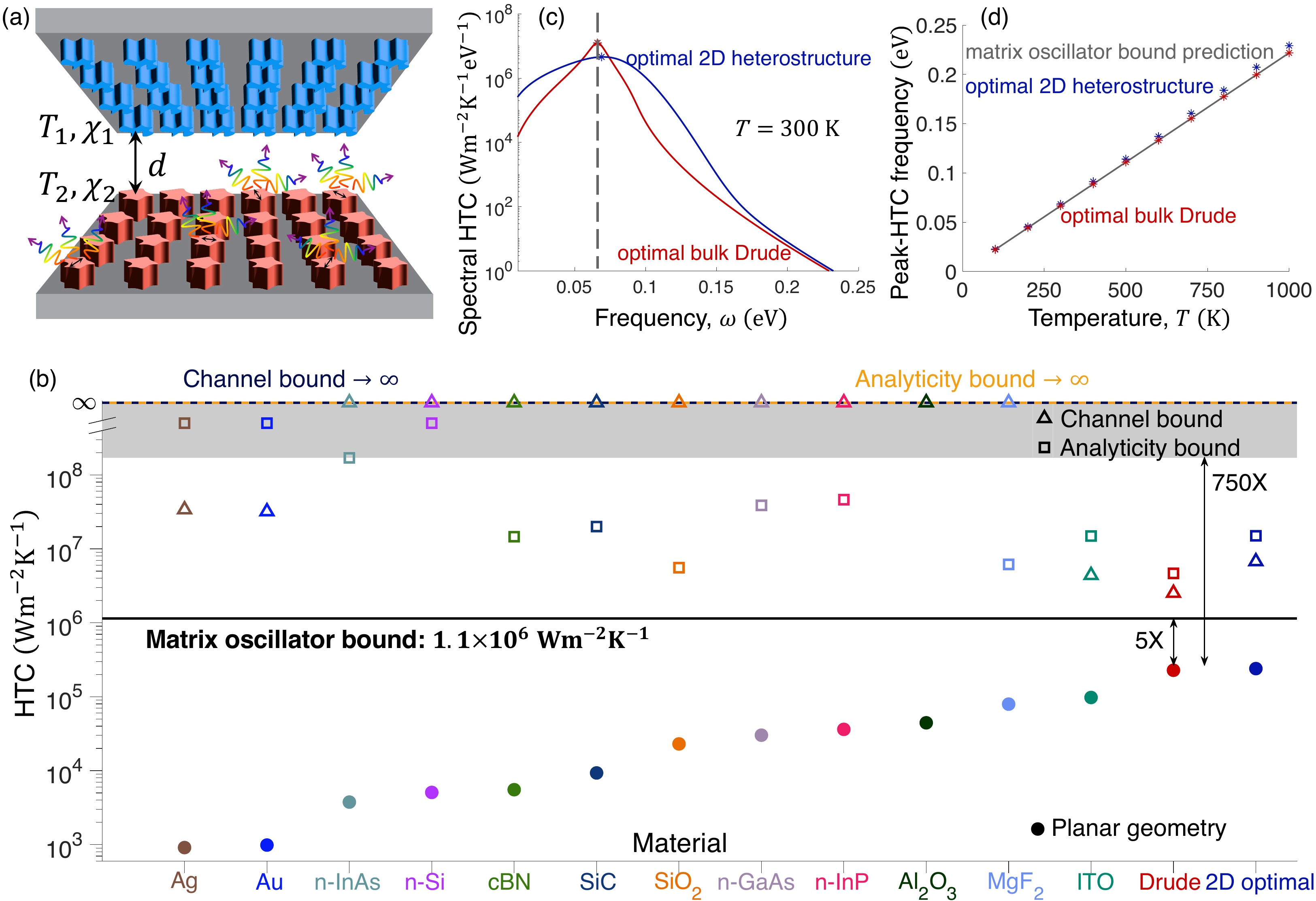}
  \caption{(a) NFRHT between two closely separated bodies. (b) Heat-transfer coefficients of planar bodies comprising increasingly high-performance planar-body geometries (filled circles), in comparison with the best previous theoretical bounds~\cite{Shim2019bandwidth,limited_role_structuring} (open squares and triangles). The previous bounds diverged for some materials, while showing enormous gaps ($>$ 750X) for others, and their trendline seems to decrease left-to-right, whereas planar-body performance increases. In black is the new theoretical bound offered by our $\TT$-matrix representation, very close to the best possible planar bodies. Panels (c,d) confirm the predictions of our spectral representation, showing that the state-of-the-art performance could be achieved precisely at the optimal frequencies identified by our approach.}
  \label{fig:NFRHT_bound}
\end{figure*}
Now we apply our formulation to the question of maximal NFRHT. It has long been known that bringing two bodies closer than a characteristic thermal wavelength can lead to radiative transfer at rates far beyond the blackbody~\cite{Polder1971,Rytov1988,Volokitin2007}, via near-field evanescent tunneling. Yet, over all possible geometrical configurations, what is the maximum possible rate of energy transfer? NFRHT, as depicted in \figref{NFRHT_bound}(a), faces prohibitive computational challenges---spatially and temporally incoherent, broadband thermal sources, exciting rapidly decaying near fields over large macroscopic areas---which has limited previous design efforts primarily to high-symmetry structures such as planar bodies~\cite{Joulain2005,Basu2009,Biehs2010}. Numerous approaches have identified particular constraints with corresponding theoretical bounds~\cite{Pendry1999,Ben-Abdallah2010,Miller2015,Shim2019bandwidth,limited_role_structuring}, but as we show in \figref{NFRHT_bound}, there are orders-of-magnitude differences between the best structures and the best bounds~\cite{Shim2019bandwidth,limited_role_structuring}. We label the bounds by their distinguishing attributes: in \citeasnoun{Shim2019bandwidth} (``analyticity bound''), complex-analyticity played the central role for a finite bound, while in \citeasnoun{limited_role_structuring} (``channel bound''), a decomposition into power-carrying channels was the starting point. Last year, it was discovered that a set of unconventional plasmonic materials offer significant (10X) improvements over the previous best planar structures~\cite{NFRHT_planar}, but otherwise the field has been at an impasse: there has been no way to identify neither better structures nor a better theory of the upper limits.

The $\TT$ matrix formulation resolves this impasse. The heat transfer coefficient (HTC) between two bodies is the net flux rate (per area and per degree $K$) of electromagnetic energy passing between bodies at temperatures $T$ and $T+\Delta T$, as measured by the integral of $(1/2) \Re \left(\Ev \times \Hv^* \cdot \hat{\vect{n}} \right)$ through a separating plane with normal vector $\hat{\vect{n}}$. The incoherent sources in body $i$ with temperature $T_i$ and susceptibility $\chi_i(\omega)$, by the fluctuation-dissipation theorem~\cite{Joulain2005}, are given by $\left\langle J_j(\xv,\omega) J_k^*(\xv',\omega) \right\rangle = (4\varepsilon_0 \omega/\pi) \Theta(\omega,T_i) \Im \left[\chi_i(\omega)\right] \delta_{jk} \delta(\xv-\xv')$ at frequency $\omega$, where $\Theta(\omega,T_i)$ is the Planck spectrum. There are a variety of mathematical transformations that we make to this problem to make it more amenable to optimization, detailed in the {\SM}, such as using reciprocity to move the sources out of the hotter body and onto the dividing surface, exploiting spatial symmetries of the bounding domains (two halfspaces, allowing for any patterning within), as well as a near-field generalization of the ``optical theorem''~\cite{Newton1976}. The key novelty, however, is our use of \eqref{OscTT}: once we have transformed the problem to an appropriate function of the two-body $\TT$ matrix, we insert the representation theorem of $\TT$ as a sum of (unknown) positive-semidefinite matrices with Drude--Lorentz lineshapes. NFRHT at moderate or high temperatures is dominated by low-frequency response (relative to the plasma frequency of common materials), which is proportional to an electrostatic constant $\alpha$ instead of the electron-density encoded in $\omega_p$. Use of the low-frequency sum rules simply requires replacement of $\omega_p^2$ with $\omega_i^2 \alpha$ in the representation theorem; moreover, the electrostatic constant is bounded above by the numerical value of 2 for high-contrast dielectric or metallic bodies enclosed in two halfspaces ({\SM}). Once we insert the $\TT$-matrix representation into the NFRHT expression, the resulting optimization problem over the infinite set of matrix oscillator coefficients has an \emph{analytical} upper bound. Straightforward algebraic manipulations ({\SM}) lead to an ultimate limit to near-field radiative HTC given by
\begin{align}
    \textrm{HTC} \leq \beta \frac{T}{d^2},
    \label{eq:HTCBound}
\end{align}
where $d$ is the minimum separation between the bodies, $T$ the temperature of the cooler body, and $\beta=0.11 (\alpha k_B^2/\hbar) = \SI{3.8e5}{W nm^2/m^2/K^2}$, a numerical constant. This limit cannot be surpassed by any geometric patterning, nor can exotic optical properties of any material alter its value.

\figref{NFRHT_bound}(b) compares our theoretical limit with the current state-of-the-art, as well as the best known bounds. Whereas the gap between the optimal planar structures and the best previous bounds was at least 750X (but diverging to $\infty$ for some materials), the expression of \eqref{HTCBound} is only 5X larger the best design. The new bound has no material dependence, which resolves the problematic trend that if one orders the materials by their planar performance, as in \figref{NFRHT_bound}(b), the previous bounds tended to predict \emph{worse} maximal performance in the same direction. The resolution of this discrepancy is our use of the low-frequency sum rule, which encodes a constraint on the local density of states seen by thermal emitters that depends only on their gap separation, independent of material. Although it seemed possible that nano-structuring may lead to enhanced NFRHT through field-concentration (lightning-rod) effects, our sum rule explains why this is not the case: sharp tips can enhance the fields very close to a sharp tip, but \emph{not} at the source location itself. The local density of states is proportional to the latter, and hence is not enhanced by lightning-rod effects. SM Sect. X illustrates this point, showing enormous field enhancements at points closer to a sharp tip than any source, but, surprisingly, smaller fields at the source itself, no matter how close the source is to the tip. The $\TT$-matrix approach predicts an optimal NFRHT frequency of $\omega_{\rm max} = 2.57\frac{k_{\rm B} T}{\hbar}$, determined by the overlap of the Planck function with the Drude--Lorentz lineshape. The predictions are matched almost exactly by computationally optimized planar Drude metals or 2D heterostructures, as shown in \figref{NFRHT_bound}(c,d). For \SI{300}{K} temperature, the spectra shown in \figref{NFRHT_bound}(c) peak at almost exactly the optimal oscillator frequency, and the match persists across all relevant temperatures, as shown in \figref{NFRHT_bound}(d).

The closeness of the arbitrary-structure bound of \eqref{HTCBound} to the best planar structures arises despite quite different mathematical routes to these results. The translational symmetry of planar bodies implies conserved wavevectors and thus a set of evanescent plane-wave channels that are independent, with Landauer-like transmissivities~\cite{Biehs2010}. Such an approach cannot describe patterned structures, and \eqref{HTCBound} culminates after using (generalized) reciprocity to move the sources from the hot body to the dividing surface, the sum rule to encapsulate the maximum densities of states seen by those sources, and the $\TT$-matrix representation to constrain the possible scattering lineshapes. The striking similarity of the two results suggests that even when confronted by spatial and temporal incoherence, rapidly decaying fields, and large areas, the oscillator representation compactly captures the key physics of maximal response in the near field.

\section{A general framework}
In this article we have introduced a new viewpoint on electromagnetic scattering. The example of \figref{Tscat} reveals a surprising structure that has previously been hidden underneath the complexity of scattering dynamics. Our application of this framework to NFRHT offers clear guidance for the fundamental limits of radiative heat transfer and the physical mechanisms underlying them. The generality of our $\TT$ matrix representation offers tantalizing prospects for wide-ranging applications across nanophotonics. Metasurfaces~\cite{Yu2014,Aieta2015}, for example, offer a new, compact form factor for optics. A central question is the extent to which metasurfaces can control incoming waves~\cite{Presutti2020,Chung2020,Engelberg2021}, across varying frequency and angular bandwidths, for applications from lenses to virtual and augmented reality. Similarly, techniques for imaging through opaque media have flourished with modern spatial light modulators~\cite{Cao2022}, with a key open question being ultimate limits to spectral control. In photovoltaics as well as photodetection, the quest for ever-thinner devices must ultimately contend with fundamental limits. And similarly across almost every application of nanophotonics. There has long been a need to quantify the ultimate limits to spectral control; our approach offers a theory to do so.

Our approach also dovetails seamlessly with a recent flurry of activity in understanding the limits controlling \emph{spatial} degrees of freedom in nanophotonic systems~\cite{Kuang2020_computational,Molesky2020,Gustafsson2020,Chao2022}. Transforming the typical Maxwell differential equations into a set of local conservation laws in space, for real and reactive power flows, leads to a mathematical form of the design problem that is amenable to systematic approaches to computational bounds. For a single frequency (or a small number of them~\cite{Shim2021}), such conservation laws have shown powerful capabilities for identifying fundamental limits to spatial control. In these approaches, the degrees of freedom of the system are typically encoded not in the electric and magnetic fields, but rather in the electric and magnetic polarization currents that they induce.  Those polarization fields are exactly those that are determined by the $\TT$ matrix, which means that our spectral expansion of the $\TT$ matrix should be seamlessly compatible with the spatial conservation laws proposed in \citeasnoun{Kuang2020_computational,Molesky2020}. Together, the two approaches may enable a complete understanding of the spatio-spectral limits of electromagnetic systems.

More broadly, the insight at the foundation of our framework, about the mathematical properties of scattering $\TT$ matrices, can be directly applied to \emph{any} classical wave equation. These techniques should be readily extensible to linear scattering problems in acoustics, elasticity, fluid dynamics, and beyond. The mathematical structure of the wave equation is similar in each case, and the resulting $\TT$ matrices should therefore have similar representations. An interesting twist may arise in acoustic scattering theory, where materials with higher-than-vacuum speeds of sound lead to ``non-causal'' scattering processes~\cite{Norris2015} that have prevented the development of classical sum rules, and would appear to prohibit a corresponding $\TT$ matrix representation. Yet the $\TT$ matrix itself may offer a new route to complex-analytic response functions in exactly such scenarios. The reason higher sound speeds lead to ``non-causal'' response is that the scattered field appears at a location within the scatterer earlier than the incident wave itself. Hence, locally, the process appears non-causal. Yet the \emph{nonlocal} nature of the $\TT$ matrix may be precisely what is needed to resolve this paradox. A $\TT$ matrix isolates the response at any point $\xv$ to the contributions from the wave incident at each point $\xv'$ in the scatterer; each of which, individually, must be causal. Hence, not only should the $\TT$ matrix be extensible to such scenarios; it may further resolve impediments that had previously stymied even simple sum rules in these fields.

Finally, we speculate that the approach described here may even be extensible to quantum scattering. In the frequency domain, the key difference between quantum and classical scattering is the analytic structure of the governing equations. In classical wave equations, second derivatives in space are proportional to second derivatives in time, which lead to poles in the lower half of the complex-frequency plane and analyticity in the upper half. In quantum scattering, second derivatives in space are proportional to first derivatives in time, which leads to bounds states for negative real energies and branch cuts on the positive real axis. Our standard semicircular contours likely need to be replaced by ``keyhole'' contours~\cite{Nussenzveig1972}, with the open question of whether there are meaningful sum rules that can be derived (perhaps dependent on bound-state properties, as in Levinson's theorem~\cite{Levinson1959,Newton2013} for spherically symmetry potentials). If such sum rules could be derived, it is likely that an infinite-oscillator description could be used to identify fundamental limits for quantum scattering as well.

\section{Acknowledgements}
L. Zhang and O.D. Miller were supported by the Air Force Office of Scientific Research Grant No. FA9550-22-1-0393 (general $\TT$-matrix theory) and by the Army Research Office Grant No. W911NF-19-1-0279 (near-field heat transfer analysis). F.M. acknowledges support from the Air Force Office of Scientific Research with Grant No. FA9550-22-1-0204, and the Office of Naval Research with Grant No. N00014-22-1-2486.

\bibliographystyle{ieeetr}
\bibliography{tmatrix}

\begin{thebibliography}{10}

\bibitem{Condon1951}
E.~U. Condon, E.~U. Condon, and G.~H. Shortley, {\em The Theory of Atomic
  Spectra}.
\newblock Cambridge University Press, 1951.

\bibitem{Bernath2005}
P.~F. Bernath, {\em Spectra of Atoms and Molecules}.
\newblock Oxford University Press, USA, Apr. 2005.

\bibitem{Shannon1948}
C.~E. Shannon, ``A mathematical theory of communication,'' {\em The Bell System
  Technical Journal}, vol.~27, pp.~379--423, July 1948.

\bibitem{Shockley1961}
W.~Shockley and H.~J. Queisser, ``Detailed balance limit of efficiency of p‐n
  junction solar cells,'' {\em J. Appl. Phys.}, vol.~32, pp.~510--519, Mar.
  1961.

\bibitem{Ching1998}
E.~S.~C. Ching, P.~T. Leung, A.~Maassen van~den Brink, W.~M. Suen, S.~S. Tong,
  and K.~Young, ``Quasinormal-mode expansion for waves in open systems,'' {\em
  Rev. Mod. Phys.}, vol.~70, pp.~1545--1554, Oct. 1998.

\bibitem{Tang2000}
S.-H. Tang and M.~Zworski, ``Resonance expansions of scattered waves,'' {\em
  Commun. Pure Appl. Math.}, vol.~LIII, pp.~1305--1334, 2000.

\bibitem{Fan_tcmt}
W.~Suh, Z.~Wang, and S.~Fan, ``Temporal coupled-mode theory and the presence of
  non-orthogonal modes in lossless multimode cavities,'' {\em IEEE Journal of
  Quantum Electronics}, vol.~40, no.~10, pp.~1511--1518, 2004.

\bibitem{Sauvan2022}
C.~Sauvan, T.~Wu, R.~Zarouf, E.~A. Muljarov, and P.~Lalanne, ``Normalization,
  orthogonality, and completeness of quasinormal modes of open systems: the
  case of electromagnetism [invited],'' {\em Opt. Express}, vol.~30,
  pp.~6846--6885, Feb. 2022.

\bibitem{Carminati2021}
R.~Carminati and J.~C. Schotland, {\em Principles of Scattering and Transport
  of Light}.
\newblock Cambridge University Press, July 2021.

\bibitem{Polder1971}
D.~Polder and M.~Van~Hove, ``Theory of radiative heat transfer between closely
  spaced bodies,'' {\em Phys. Rev. B: Condens. Matter Mater. Phys.}, vol.~4,
  no.~10, pp.~3303--3314, 1971.

\bibitem{Rytov1988}
S.~M. Rytov, Y.~A. Kravtsov, and V.~I. Tatarskii, {\em Principles of
  Statistical Radiophysics}.
\newblock New York, NY: Springer-Verlag, 1988.

\bibitem{Volokitin2007}
A.~I. Volokitin and B.~N.~J. Persson, ``Near-field radiative heat transfer and
  noncontact friction,'' {\em Rev. Mod. Phys.}, vol.~79, no.~4, pp.~1291--1329,
  2007.

\bibitem{Fiorino2018}
A.~Fiorino, L.~Zhu, D.~Thompson, R.~Mittapally, P.~Reddy, and E.~Meyhofer,
  ``Nanogap near-field thermophotovoltaics,'' {\em Nature nanotechnology},
  vol.~13, no.~9, pp.~806--811, 2018.

\bibitem{Bhatt2020}
G.~R. Bhatt, B.~Zhao, S.~Roberts, I.~Datta, A.~Mohanty, T.~Lin, J.-M. Hartmann,
  R.~St-Gelais, S.~Fan, and M.~Lipson, ``Integrated near-field
  thermo-photovoltaics for heat recycling,'' {\em Nature communications},
  vol.~11, no.~1, pp.~1--7, 2020.

\bibitem{Zhu2019}
L.~Zhu, A.~Fiorino, D.~Thompson, R.~Mittapally, E.~Meyhofer, and P.~Reddy,
  ``Near-field photonic cooling through control of the chemical potential of
  photons,'' {\em Nature}, vol.~566, no.~7743, pp.~239--244, 2019.

\bibitem{Challener2009}
W.~Challener, C.~Peng, A.~Itagi, D.~Karns, W.~Peng, Y.~Peng, X.~Yang, X.~Zhu,
  N.~Gokemeijer, Y.-T. Hsia, {\em et~al.}, ``Heat-assisted magnetic recording
  by a near-field transducer with efficient optical energy transfer,'' {\em
  Nature photonics}, vol.~3, no.~4, pp.~220--224, 2009.

\bibitem{Pendry1999}
J.~B. Pendry, ``Radiative exchange of heat between nanostructures,'' {\em J.
  Phys. Condens. Matter}, vol.~11, no.~35, pp.~6621--6633, 1999.

\bibitem{Ben-Abdallah2010}
P.~Ben-Abdallah and K.~Joulain, ``Fundamental limits for noncontact transfers
  between two bodies,'' {\em Phys. Rev. B: Condens. Matter Mater. Phys.},
  vol.~82, p.~121419, 2010.

\bibitem{Miller2015}
O.~D. Miller, S.~G. Johnson, and A.~W. Rodriguez, ``Shape-independent limits to
  near-field radiative heat transfer,'' {\em Physical Review Letters},
  vol.~115, p.~204302, Nov. 2015.

\bibitem{Shim2019bandwidth}
H.~Shim, L.~Fan, S.~G. Johnson, and O.~D. Miller, ``Fundamental limits to
  near-field optical response over any bandwidth,'' {\em Phys. Rev. X}, vol.~9,
  p.~011043, Mar 2019.

\bibitem{limited_role_structuring}
P.~S. Venkataram, S.~Molesky, W.~Jin, and A.~W. Rodriguez, ``Fundamental limits
  to radiative heat transfer: The limited role of nanostructuring in the
  near-field,'' {\em Phys. Rev. Lett.}, vol.~124, p.~013904, Jan 2020.

\bibitem{Burke2012}
K.~Burke, ``Perspective on density functional theory,'' {\em The Journal of
  chemical physics}, vol.~136, no.~15, p.~150901, 2012.

\bibitem{Jackson1999}
J.~D. Jackson, {\em Classical Electrodynamics, 3rd Ed}.
\newblock John Wiley \& Sons, 1999.

\bibitem{Andreoli2021}
F.~Andreoli, M.~J. Gullans, A.~A. High, A.~Browaeys, and D.~E. Chang, ``Maximum
  refractive index of an atomic medium,'' {\em Phys. Rev. X}, vol.~11,
  p.~011026, Feb. 2021.

\bibitem{Shim_refractive_index}
H.~Shim, F.~Monticone, and O.~D. Miller, ``Fundamental limits to the refractive
  index of transparent optical materials,'' {\em Advanced Materials}, vol.~33,
  no.~43, p.~2103946, 2021.

\bibitem{Martinsson2019}
P.~G. Martinsson, {\em Fast Direct Solvers for Elliptic {PDEs}}.
\newblock Philadelphia, PA: Society for Industrial and Applied Mathematics,
  Jan. 2019.

\bibitem{Nussenzveig1972}
H.~M. Nussenzveig, {\em Causality and Dispersion Relations}.
\newblock New York, NY: Academic Press, 1972.

\bibitem{Reiche1925}
F.~Reiche and W.~Thomas, ``{{\"{U}}ber die Zahl der Dispersionselektronen, die
  einem station{\"{a}}ren Zustande zugeordnet sind. (Vorl{\"{a}}ufige
  Mitteilung)},'' {\em Zeitschrift f{\"{u}}r Physik}, vol.~34, pp.~510--525,
  jul 1925.

\bibitem{Kuhn1925}
W.~Kuhn, ``{{\"{U}}ber die Gesamtst{\"{a}}rke der von einem Zustande
  ausgehenden Absorptionslinien},'' {\em Zeitschrift f{\"{u}}r Physik},
  vol.~33, pp.~408--412, dec 1925.

\bibitem{sum_rule_chi}
F.~W. King, ``Sum rules for the optical constants,'' {\em Journal of
  Mathematical Physics}, vol.~17, no.~8, pp.~1509--1514, 1976.

\bibitem{Joulain2005}
K.~Joulain, J.-P. Mulet, F.~Marquier, R.~Carminati, and J.-J. Greffet,
  ``Surface electromagnetic waves thermally excited: Radiative heat transfer,
  coherence properties and casimir forces revisited in the near field,'' {\em
  Surf. Sci. Rep.}, vol.~57, pp.~59--112, May 2005.

\bibitem{Basu2009}
S.~Basu, Z.~M. Zhang, and C.~J. Fu, ``Review of near-field thermal radiation
  and its application to energy conversion,'' {\em Int. J. Energy Res.},
  vol.~33, pp.~1203--1232, 2009.

\bibitem{Biehs2010}
S.-A. Biehs, E.~Rousseau, and J.-J. Greffet, ``Mesoscopic description of
  radiative heat transfer at the nanoscale,'' {\em Phys. Rev. Lett.}, vol.~105,
  p.~234301, Dec. 2010.

\bibitem{NFRHT_planar}
L.~Zhang and O.~D. Miller, ``Optimal materials for maximum large-area
  near-field radiative heat transfer,'' {\em ACS Photonics}, vol.~7, no.~11,
  pp.~3116--3129, 2020.

\bibitem{Newton1976}
R.~G. Newton, ``Optical theorem and beyond,'' {\em Am. J. Phys.}, vol.~44,
  no.~7, pp.~639--642, 1976.

\bibitem{Yu2014}
N.~Yu and F.~Capasso, ``Flat optics with designer metasurfaces,'' {\em Nat.
  Mater.}, vol.~13, no.~2, pp.~139--150, 2014.

\bibitem{Aieta2015}
F.~Aieta, M.~A. Kats, P.~Genevet, and F.~Capasso, ``Multiwavelength achromatic
  metasurfaces by dispersive phase compensation,'' {\em Science}, vol.~347,
  no.~6228, pp.~1342--1345, 2015.

\bibitem{Presutti2020}
F.~Presutti and F.~Monticone, ``Focusing on bandwidth: achromatic metalens
  limits,'' {\em Optica}, vol.~7, p.~624, June 2020.

\bibitem{Chung2020}
H.~Chung and O.~D. Miller, ``High-{NA} achromatic metalenses by inverse
  design,'' {\em Opt. Express}, vol.~28, pp.~6945--6965, Feb. 2020.

\bibitem{Engelberg2021}
J.~Engelberg and U.~Levy, ``Achromatic flat lens performance limits,'' {\em
  Optica}, vol.~8, p.~834, June 2021.

\bibitem{Cao2022}
H.~Cao, A.~P. Mosk, and S.~Rotter, ``Shaping the propagation of light in
  complex media,'' {\em Nat. Phys.}, vol.~18, pp.~994--1007, Sept. 2022.

\bibitem{Kuang2020_computational}
Z.~Kuang and O.~D. Miller, ``Computational bounds to light--matter interactions
  via local conservation laws,'' {\em Phys. Rev. Lett.}, vol.~125, p.~263607,
  Dec 2020.

\bibitem{Molesky2020}
S.~Molesky, P.~Chao, and A.~W. Rodriguez, ``Hierarchical mean-field {T}
  operator bounds on electromagnetic scattering: Upper bounds on near-field
  radiative purcell enhancement,'' {\em Phys. Rev. Research}, vol.~2,
  p.~043398, Dec. 2020.

\bibitem{Gustafsson2020}
M.~Gustafsson, K.~Schab, L.~Jelinek, and M.~Capek, ``Upper bounds on absorption
  and scattering,'' {\em New Journal of Physics}, vol.~22, p.~073013, 2020.

\bibitem{Chao2022}
P.~Chao, B.~Strekha, R.~Kuate~Defo, S.~Molesky, and A.~W. Rodriguez, ``Physical
  limits in electromagnetism,'' {\em Nat. Rev. Phys.}, vol.~4, pp.~543--559,
  2022.

\bibitem{Shim2021}
H.~Shim, Z.~Kuang, Z.~Lin, and O.~D. Miller, ``Fundamental limits to
  multi-functional and tunable nanophotonic response,'' Dec. 2021.

\bibitem{Norris2015}
A.~N. Norris, ``Acoustic integrated extinction,'' {\em Proceedings of the Royal
  Society A: Mathematical, Physical and Engineering Sciences}, vol.~471,
  no.~2177, p.~20150008, 2015.

\bibitem{Levinson1959}
N.~Levinson, ``Kgl. danske videnskab selskab mat.-fys. medd. 25 (1949), no. 9.
  m. ida,'' {\em Prog. Theor. Phys}, vol.~21, p.~625, 1959.

\bibitem{Newton2013}
R.~G. Newton, {\em Scattering theory of waves and particles}.
\newblock Springer Science \& Business Media, 2013.

\end{thebibliography}


\begin{thebibliography}{10}

\bibitem{Suh2004}
W.~Suh, Z.~Wang, and S.~Fan, ``Temporal coupled-mode theory and the presence of
  non-orthogonal modes in lossless multimode cavities,'' {\em IEEE J. Quantum
  Electron.}, vol.~40, pp.~1511--1518, Oct. 2004.

\bibitem{Haus1984}
H.~A. Haus, {\em Waves and fields in optoelectronics}.
\newblock Prentice-Hall, 1984.

\bibitem{Joannopoulos2011}
J.~D. Joannopoulos, S.~G. Johnson, J.~N. Winn, and R.~D. Meade, {\em Photonic
  crystals: molding the flow of light}.
\newblock Princeton University Press, 2011.

\bibitem{Fan2003}
S.~Fan, W.~Suh, and J.~D. Joannopoulos, ``Temporal coupled-mode theory for the
  fano resonance in optical resonators,'' {\em J. Opt. Soc. Am.}, vol.~20,
  no.~3, p.~569, 2003.

\bibitem{Hamam2007}
R.~E. Hamam, A.~Karalis, J.~D. Joannopoulos, and M.~Solja{\v c}i{\'c},
  ``Coupled-mode theory for general free-space resonant scattering of waves,''
  {\em Phys. Rev. A}, vol.~75, p.~053801, May 2007.

\bibitem{Ching1998}
E.~S.~C. Ching, P.~T. Leung, A.~Maassen van~den Brink, W.~M. Suen, S.~S. Tong,
  and K.~Young, ``Quasinormal-mode expansion for waves in open systems,'' {\em
  Rev. Mod. Phys.}, vol.~70, pp.~1545--1554, Oct. 1998.

\bibitem{lalanne2018light}
P.~Lalanne, W.~Yan, K.~Vynck, C.~Sauvan, and J.-P. Hugonin, ``Light interaction
  with photonic and plasmonic resonances,'' {\em Laser \& Photonics Reviews},
  vol.~12, no.~5, p.~1700113, 2018.

\bibitem{sauvan2013theory}
C.~Sauvan, J.-P. Hugonin, I.~Maksymov, and P.~Lalanne, ``Theory of the
  spontaneous optical emission of nanosize photonic and plasmon resonators,''
  {\em Physical Review Letters}, vol.~110, no.~23, p.~237401, 2013.

\bibitem{muljarov2016resonant}
E.~Muljarov and W.~Langbein, ``Resonant-state expansion of dispersive open
  optical systems: Creating gold from sand,'' {\em Physical Review B}, vol.~93,
  no.~7, p.~075417, 2016.

\bibitem{lalanne2019quasinormal}
P.~Lalanne, W.~Yan, A.~Gras, C.~Sauvan, J.-P. Hugonin, M.~Besbes,
  G.~Dem{\'e}sy, M.~Truong, B.~Gralak, F.~Zolla, {\em et~al.}, ``Quasinormal
  mode solvers for resonators with dispersive materials,'' {\em JOSA A},
  vol.~36, no.~4, pp.~686--704, 2019.

\bibitem{kristensen2017theory}
P.~T. Kristensen, J.~R. de~Lasson, M.~Heuck, N.~Gregersen, and J.~M{\o}rk, ``On
  the theory of coupled modes in optical cavity-waveguide structures,'' {\em
  Journal of Lightwave Technology}, vol.~35, no.~19, pp.~4247--4259, 2017.

\bibitem{kristensen2020modeling}
P.~T. Kristensen, K.~Herrmann, F.~Intravaia, and K.~Busch, ``Modeling
  electromagnetic resonators using quasinormal modes,'' {\em Advances in Optics
  and Photonics}, vol.~12, no.~3, pp.~612--708, 2020.

\bibitem{Zhang2020}
H.~Zhang and O.~D. Miller, ``Quasinormal coupled mode theory,'' {\em arXiv
  preprint arXiv:2010.08650}, 2020.

\bibitem{mahaux1969shell}
C.~Mahaux and H.~A. Weidenm{\"u}ller, {\em Shell-model approach to nuclear
  reactions.}
\newblock North-Holland Pub. Co., 1969.

\bibitem{sweeney2019theory}
W.~R. Sweeney, C.~W. Hsu, and A.~D. Stone, ``Theory of reflectionless
  scattering modes,'' {\em arXiv preprint arXiv:1909.04017}, 2019.

\bibitem{Miller2017}
D.~A.~B. Miller, L.~Zhu, and S.~Fan, ``Universal modal radiation laws for all
  thermal emitters,'' {\em Proc. Natl. Acad. Sci. U. S. A.}, vol.~114, no.~17,
  pp.~4336--4341, 2017.

\bibitem{Yu2010}
Z.~Yu, A.~Raman, and S.~Fan, ``Fundamental limit of nanophotonic light trapping
  in solar cells,'' {\em Proc. Natl. Acad. Sci. U. S. A.}, vol.~107,
  pp.~17491--17496, Oct. 2010.

\bibitem{Yu2011}
Z.~Yu, A.~Raman, and S.~Fan, ``Nanophotonic light-trapping theory for solar
  cells,'' {\em Appl. Phys. A: Mater. Sci. Process.}, vol.~105, pp.~329--339,
  Nov. 2011.

\bibitem{Yu2012}
Z.~Yu, A.~Raman, and S.~Fan, ``Thermodynamic upper bound on broadband light
  coupling with photonic structures,'' {\em Phys. Rev. Lett.}, vol.~109,
  p.~173901, Oct. 2012.

\bibitem{Verslegers2010}
L.~Verslegers, Z.~Yu, P.~B. Catrysse, and S.~Fan, ``Temporal coupled-mode
  theory for resonant apertures,'' {\em J. Opt. Soc. Am. B}, vol.~27, p.~1947,
  Oct. 2010.

\bibitem{Verslegers2012}
L.~Verslegers, Z.~Yu, Z.~Ruan, P.~B. Catrysse, and S.~Fan, ``From
  electromagnetically induced transparency to superscattering with a single
  structure: A {Coupled-Mode} theory for doubly resonant structures,'' {\em
  Phys. Rev. Lett.}, vol.~108, p.~83902, Feb. 2012.

\bibitem{Hsu2014}
C.~W. Hsu, B.~G. DeLacy, S.~G. Johnson, J.~D. Joannopoulos, and M.~Solja{\v
  c}i{\'c}, ``Theoretical criteria for scattering dark states in nanostructured
  particles,'' {\em Nano Lett.}, vol.~14, pp.~2783--2788, May 2014.

\bibitem{Zhou2016}
H.~Zhou, B.~Zhen, C.~W. Hsu, O.~D. Miller, S.~G. Johnson, J.~D. Joannopoulos,
  and M.~Solja{\v c}i{\'c}, ``Perfect single-sided radiation and absorption
  without mirrors,'' {\em Optica}, vol.~3, pp.~1079--1086, Oct. 2016.

\bibitem{Hsu2016}
C.~W. Hsu, B.~Zhen, A.~D. Stone, J.~D. Joannopoulos, and M.~Solja{\v c}i{\'c},
  ``Bound states in the continuum,'' {\em Nature Reviews Materials}, vol.~1,
  pp.~1--13, July 2016.

\bibitem{Mann2019}
S.~A. Mann, D.~L. Sounas, and A.~Al{\`u}, ``Nonreciprocal cavities and the
  time--bandwidth limit,'' {\em Optica}, vol.~6, pp.~104--110, Jan. 2019.

\bibitem{Benzaouia2021}
M.~Benzaouia, J.~D. Joannopoulos, S.~G. Johnson, and A.~Karalis, ``Quasi-normal
  mode theory of the scattering matrix, enforcing fundamental constraints for
  truncated expansions,'' {\em Phys. Rev. Research}, vol.~3, p.~033228, Sept.
  2021.

\bibitem{Chew2008}
W.~C. Chew, M.~S. Tong, and B.~Hu, ``Integral equation methods for
  electromagnetic and elastic waves,'' {\em Synthesis Lectures on Computational
  Electromagnetics}, vol.~3, pp.~1--241, 2008.

\bibitem{Nussenzveig1972}
H.~M. Nussenzveig, {\em Causality and Dispersion Relations}.
\newblock New York, NY: Academic Press, 1972.

\bibitem{Lucarini2005}
V.~Lucarini, J.~J. Saarinen, K.-E. Peiponen, and E.~M. Vartiainen, {\em
  {Kramers-Kronig} Relations in Optical Materials Research}.
\newblock Springer Science \& Business Media, Apr. 2005.

\bibitem{static_Fresnel}
J.~Skaar, ``Fresnel's equations in statics and quasistatics,'' {\em European
  Journal of Physics}, vol.~40, p.~045201, Jun. 2019.

\bibitem{Ivanenko2019}
Y.~Ivanenko, M.~Gustafsson, B.~L.~G. Jonsson, A.~Luger, B.~Nilsson, S.~Nordebo,
  and J.~Toft, ``Passive approximation and optimization using b-splines,'' {\em
  SIAM J. Appl. Math.}, vol.~79, pp.~436--458, Jan. 2019.

\bibitem{King2009}
F.~W. King, {\em Hilbert Transforms: Volume 2}, vol.~2.
\newblock Cambridge University Press, 2009.

\bibitem{Bernland2011}
A.~Bernland, A.~Luger, and M.~Gustafsson, ``Sum rules and constraints on
  passive systems,'' {\em J. Phys. A: Math. Theor.}, vol.~44, p.~145205, Mar.
  2011.

\bibitem{Luger2022}
A.~Luger and M.-J.~Y. Ou, ``On applications of {Herglotz-Nevanlinna} functions
  in material sciences, {I}: classical theory and applications of sum rules,''
  {\em arXiv:2202.13247}, Feb. 2022.

\bibitem{Ou2022}
M.-J.~Y. Ou and A.~Luger, ``On applications of {Herglotz-Nevanlinna} functions
  in material sciences, {II}: extended applications and generalized theory,''
  {\em arXiv:2202.13246}, Feb. 2022.

\bibitem{King2006}
F.~W. King, ``Alternative approach to the derivation of dispersion relations
  for optical constants,'' {\em J. Phys. A Math. Gen.}, vol.~39, p.~10427, Aug.
  2006.

\bibitem{Gesztesy2000}
F.~Gesztesy and E.~Tsekanovskii, ``On matrix-valued herglotz functions,'' {\em
  Math. Nachr.}, vol.~218, pp.~61--138, Oct. 2000.

\bibitem{Fritzsche2012}
B.~Fritzsche, B.~Kirstein, and C.~M{\"a}dler, ``On matrix-valued
  {Herglotz-Nevanlinna} functions with an emphasis on particular subclasses,''
  {\em Math. Nachr.}, vol.~285, pp.~1770--1790, Oct. 2012.

\bibitem{Yurkin2011DDA}
M.~A. Yurkin and A.~G. Hoekstra, ``The discrete-dipole-approximation code adda:
  Capabilities and known limitations,'' {\em Journal of Quantitative
  Spectroscopy and Radiative Transfer}, vol.~112, pp.~2234--2247, Sept. 2011.

\bibitem{Duan2009}
R.~Duan and V.~Rokhlin, ``High-order quadratures for the solution of scattering
  problems in two dimensions,'' {\em J. Comput. Phys.}, vol.~228,
  p.~2152–2174, Apr. 2009.

\bibitem{Shim2019bandwidth}
H.~Shim, L.~Fan, S.~G. Johnson, and O.~D. Miller, ``Fundamental limits to
  near-field optical response over any bandwidth,'' {\em Phys. Rev. X}, vol.~9,
  p.~011043, Mar 2019.

\bibitem{Absil2009}
P.-A. Absil, R.~Mahony, and R.~Sepulchre, {\em Optimization algorithms on
  matrix manifolds}.
\newblock Princeton University Press, 2009.

\bibitem{manopt}
N.~Boumal, B.~Mishra, P.-A. Absil, and R.~Sepulchre, ``{M}anopt, a {M}atlab
  toolbox for optimization on manifolds,'' {\em Journal of Machine Learning
  Research}, vol.~15, no.~42, pp.~1455--1459, 2014.

\bibitem{NFRHT_planar}
L.~Zhang and O.~D. Miller, ``Optimal materials for maximum large-area
  near-field radiative heat transfer,'' {\em ACS Photonics}, vol.~7, no.~11,
  pp.~3116--3129, 2020.

\end{thebibliography}

\end{document}


\title{Supplementary Materials:\\All electromagnetic scattering bodies are matrix-valued oscillators}
\author{Lang Zhang}
\author{Francesco Monticone}
\author{Owen D. Miller}

\date{\today}

\maketitle
\tableofcontents

\newpage

\section{Physical oscillators vs. mathematical oscillators}
In the main text we contrasted our $\TT$-matrix ``mathematical-oscillator'' representation with the well-known ``physical-oscillator'' decompositions in use today. Here we detail the similarities and differences in these approaches. At the highest level, physical-oscillator approaches are meant to be efficient for simulation and modeling: for a \emph{given} structure, can one identify a small number of parameters (e.g. normal- or quasinormal-mode coefficients, etc.) that accurately model the complete response of the system? Typically these models are highly nonlinear in the unknown parameters, but there are standard computational methods for finding them. Yet for problems of design, these representations are difficult or impossible to work with: there is not a single given structure, anymore, but instead a large class of structures. It is not known \emph{a priori} how many resonances or modes may contribute; to be safe, very large numbers must be used. So one is left with large, highly nonlinear models to optimize over, without any beneficial mathematical structure. The ``mathematical-oscillator'' theory developed in the main text is well-suited to this scenario. The use of lossless (and hence narrow-linewidth) oscillators naturally also leads to a large numbers of parameters (matrix-valued oscillator coefficients), but these parameters have ideal properties from an optimization perspective: they are positive-definite, constrained in sum, and linear in the only degrees of freedom. In fact, this decomposition is quite \emph{ill-suited} for modeling: one needs to invert a large, dense matrix at every frequency to get the corresponding coefficients, which is computationally prohibitive except for small structures. But for \emph{design}, one never needs to do any matrix inversion, and the mathematical structure of the decomposition is far superior for optimization. Below, we provide the mathematical expressions supporting these qualitative assertions.

In electromagnetic scattering simulations, there are two primary classes of ``physical-oscillator'' approaches: coupled-mode theory (CMT)~\cite{Suh2004,Haus1984,Joannopoulos2011,Fan2003,Hamam2007}, and quasinormal-mode (QNM) theories~\cite{Ching1998,lalanne2018light,sauvan2013theory,muljarov2016resonant,lalanne2019quasinormal,kristensen2017theory,kristensen2020modeling}. It can be shown that the former can be derived from the latter in the limit of isolated, high-$Q$ resonances with negligible non-resonant scattering contributions~\cite{Zhang2020}. We will describe both of these constructions. We start with coupled-mode theory. In coupled-mode theory, there is a basis of resonant modes described by a matrix $\Omega$, whose diagonal terms are \emph{complex-valued} resonance frequencies of the modes, and whose off-diagonal terms describe coupling rates between each pair of modes. Each resonance has ```overlap coeffients'' with each outgoing-wave channel, the matrix containing these elements is often referred to as $K$ (or $D$, which is identical to $K$ in reciprocal systems). Finally, there is typically a non-trivial background scattering matrix $S_{\rm bg}$ that contains non-resonant contributions to the scattering process. In full, in any general (reciprocal) coupled-mode theory, the scattering matrix is given by the expression~\cite{Suh2004,Zhang2020}
\begin{align}
    S = S_{\rm bg} - i K \left(\Omega - \omega\right)^{-1} K^T,
    \label{eq:SCMT}
\end{align}
subject to reciprocity and unitarity conditions given by
\begin{align}
    K^\dagger K &= 2 \Im \Omega, \\
    S_{\rm bg} K^* &= -K.
    \label{eq:SKrelation}
\end{align}
One can immediately see that a CMT model constructed from these three equations will be impossible to optimize over. The degrees of freedom are the matrices $\Omega$, $S_{\rm bg}$, and $K$, none of which are Hermitian (let alone positive definite). Moreover, one cannot even presuppose any finite, constrained size of the matrices, as there is no sum rule constraining any norm of the entries. 

One route towards using CMT models for understanding limits is to remove much of the complexity from \eqrefrange{SCMT}{SKrelation}. If one assumes background scattering cannot occur (although note that even in Mie resonators it plays a quite important role~\cite{Zhang2020}), then $S_{\rm bg} = I$ and $K^* = -K$, such that one can rewrite the scattering matrix relation as
\begin{align}
    S = \II + i K \left(\Omega - \omega\right)^{-1} K^\dagger,
\end{align}
a form of the $S$-matrix that also arises in nuclear scattering theory~\cite{mahaux1969shell,sweeney2019theory}. Next, one can assume that none of the modes are coupled, such that $\Omega$ is a diagonal matrix. (This condition will typically conflict with the requirement that $K^\dagger K = 2\Im \Omega$, but we ignore that for simplicity.) Finally, special quantities such as absorption (given by $\II - S^\dagger S$~\cite{Miller2017}) can be written as:
\begin{align}
    A = \II - S^\dagger S = -4K\left(\Omega - \omega\right)^{-\dagger} \left[\Im \Omega\right] \left(\Omega - \omega\right)^{-1} K^\dagger,
    \label{eq:ACMT}
\end{align}
after repeated use of $K^\dagger K = 2\Im \Omega$ and the matrix identity $\Im\left[ X^\dagger Y X\right] = X^\dagger \left[\Im Y\right] X$. From \eqref{ACMT}, one can integrate over all frequencies to simplify the interior matrix product involving frequencies (which is a diagonal matrix with Lorentzians along the diagonal) to a constant. Finally, one is left with an expression involving only the loss rates of each resonator and the number of resonances~\cite{Yu2010,Yu2011,Yu2012}. But what are these values? How large can they be? One is always left with more free unconstrained parameters. And the extreme limits of these models, where one wants to operate for fundamental limits, are precisely where the assumptions mentioned above (high-$Q$ resonances, uncoupled resonances, frequency-independent $K$ matrix, isolated resonances, no background processes, etc.) break down. By contrast, the utility of CMT for \emph{modeling} complex electromagnetic structures has been a theoretical bounty for twenty years~\cite{Suh2004,Joannopoulos2011,Fan2003,Hamam2007,Verslegers2010,Verslegers2012,Hsu2014,Zhou2016,Hsu2016,Mann2019}, and CMT is well-deserving of its popularity for such scenarios.

To move beyond the assumptions of CMT, expansions via quasinormal modes (QNMs) have become more popular in recent years. There are various expansion techniques, many of which can be shown to be equivalent~\cite{lalanne2018light}. We will assume a Maxwell equation of the form
\begin{align}
    (M - \omega^2 \varepsilon)e = i\omega j,
\end{align}
where we assume a sufficiently high-resolution discretization of Maxwell's equations, for matrix $M$ and diagonal matrix $\varepsilon$, unknown electric-field vector $e$, and free-current source vector $j$. The boundary conditions or PMLs are assumed to be encoded in the matrix $M$, as well as the curl--curl operator. The pair of matrices $M$ and $\varepsilon$ form generalized eigenproblem pairs according to $MU = \varepsilon U \Lambda$, where $U$ are the eigenfields and $\Lambda$ the squared eigenfrequencies. We can assume reciprocity, in which case $U^{-1} = U^T$. Inserting this eigendecomposition into our Maxwell equation and solving for the electric field yields
\begin{align}
    e = i\omega U \left(\Lambda - \omega^2 \right)^{-1} U^T \varepsilon^{-1} j.
    \label{eq:eQNM}
\end{align}
One can interpret \eqref{eQNM} intuitively: $U^T \varepsilon^{-1} j$ is a decomposition of normalized free currents into modal fields, $\left(\Lambda - \omega^2 \right)^{-1}$ is the resonant enhancement associated with real frequencies close to the resonant frequencies, and the final $U$ on the left converts from the modal basis back to the original (real-space) basis. Superficially, \eqref{eQNM} actually looks quite similar to the coupled-mode scattering-matrix equation of \eqref{SCMT}: a resonant amplification term inversely proportional to the differences between the complex-valued resonant frequencies and the real excitation frequencies, and frequency-independent matrices surrounding the resonant-amplification term. However, to connect to the ``scattering channels'' that bring energy into or out of such systems, one would need to pre- and post-multiply these matrices with Green's-function matrices that are highly frequency-dependent. Then, again, one is left with a complex set of degrees of freedom: the number of resonances (the size of $\Lambda$ and number of columns of $U$), the locations of the resonant poles in the complex plane (the values of $\Lambda$), and the resonant field patterns (the values of the columns of $U$). There are almost no constraints on these degrees of freedom, except that the field patterns must be orthogonal in the unconjugated inner product, corresponding to $U^T U = \II$. There is no meaningful way to convert this representation to upper bounds or fundamental limits. Again, however, this representation is quite useful for modeling, with a number of exemplary successes over the past decade~\cite{Ching1998,lalanne2018light,sauvan2013theory,muljarov2016resonant,lalanne2019quasinormal,kristensen2017theory,kristensen2020modeling,Benzaouia2021}.

To summarize: \eqreftwo{SCMT}{eQNM} are the key ``physical-oscillator'' descriptions of classical scattering processes. At a glance, they share similarities with each other and with the $\TT$-matrix representation of the main text: at a coarse level, each has a resonant-enhancement term and one or more matrices that can be described as a ``coupling'' matrix. With more granularity, however, there are crucial mathematical differences between the two expressions of \eqreftwo{SCMT}{eQNM} with the $\TT$ matrix expression. In the CMT and QNM approaches, all of the degrees of freedom (the resonant pole locations and the coupling matrices) are complex-valued quantities without any Hermiticity or positive-definiteness qualities. Moreover, the number of resonances can never be constrained for the arbitrarily patterned nanophotonic systems of interest. By contrast, in the $\TT$ matrix expansion, all of the ``resonant poles'' are approaching the real axis, there is an infinite set (one need not limit the number of ``resonances''), the degrees of freedom (the scattering-oscillator strengths) are positive semidefinite, Hermitian matrices, and their sum is constrained, thanks to sum rules. Because their poles are \emph{not} related to the normal- or quasinormal-mode eigenfrequencies, $\TT$ matrix expansions are computationally expensive for a given structure. But in optimizations over \emph{all possible} geometries, their mathematical structure is unique, and pays significant dividends.

\section{Integral-operator definition of the $\TT$ matrix}
In the main text, we used linearity as a sufficient condition to argue that the polarization fields $\Pv(\xv)$ induced by an incident field $\Einc(\xv)$ must be related through a linear operator that one can call ``$\TT$:''
\begin{align}
    \Pv(\xv) = \int_V \TT(\xv,\xv') \Einc(\xv') \,{\rm d}\xv',
\end{align}
or, in our vector notation,
\begin{align}
    \pv = \TT \einc.
\end{align}
In this section, we show how to \emph{construct} (or define) the $\TT$ matrix from known integral-equation operators.

For any scattering problem, one can solve for all field degrees of freedom within only the body of the scatterer, from which all other fields are computable. One starts by writing the total field as the sum of the incident and scattered fields,
\begin{align}
    \Ev(\xv) = \Einc(\xv) + \Ev_{\rm scat}(\xv),
    \label{eq:eTot}
\end{align}
where $\Einc(\xv)$ is known (for a given application). The scattered field is given by the convolution of the free-space (or background) Green's function $\GG_0$ with the polarization fields:
\begin{align}
    \Ev_{\rm scat}(\xv) = \int_V \GG_0(\xv,\xv') \Pv(\xv') \,{\rm d}\xv'.
\end{align}
Within the scatterer, the total field $\Ev(\xv)$ equals the polarization field $\Pv(\xv)$ divided by the susceptibility $\chi(\xv)$ (which is nonzero everywhere by definition of ``within the scatterer''), so that we can rewrite the decomposition of the electric field in \eqref{eTot} in terms of the polarization fields
\begin{align}
    \int_V \GG_0(\xv,\xv') \Pv(\xv') \,{\rm d}\xv' - \frac{1}{\chi(\xv)}\Pv(\xv) = -\Einc(\xv),
\end{align}
or, in vector notation,
\begin{align}
    \left[\GG_0 - \chi^{-1}\right] \pv = -\einc,
    \label{eq:LS}
\end{align}
which in electromagnetism is known as the volume integral equation, or Lippmann--Schwinger equation~\cite{Chew2008}. We can simply invert the matrices in square brackets,
\begin{align}
    \pv = -\left[\GG_0 - \chi^{-1}\right]^{-1} \einc,
\end{align}
to find the definition of $\TT$:
\begin{align}
    \TT = -\left[\GG_0 - \chi^{-1}\right]^{-1}.
    \label{eq:TTGF}
\end{align}
Hence the $\TT$ matrix is the inverse of well-known integral-equation matrices, which can be computed via standard methods~\cite{Chew2008}.

\section{$\TT$ matrix Kramers--Kronig relation: Reciprocal Case}
In this section, we fill in the details for the derivation of our ``Kramers--Kronig'' $\TT$ matrix equation, Eq. (2) of the main text. Our derivation mirrors the derivation of conventional KK relations for susceptibilities, cf.~\cite{Nussenzveig1972,Lucarini2005}.

We start with the time-domain equation defining the $\TT$ matrix as the linear matrix connecting incident fields to induced polarization fields:
\begin{align}
    \pv(t) = \int \TT(t-t') \einc(t') \,{\rm d}t'.
\end{align}
This definition enables us to encode causality in the $\TT$: at any point $\xv$ in the scatterer, the polarization field at that point cannot be generated from the incident field at any other point $\xv'$ until \emph{after} the incident field has reached that point $\xv'$. We can define the first time at which the incident field arrives at any point in the scatterer as the time origin, in which case causality implies that 
\begin{align}
    \TT(t<0) = 0,
    \label{eq:causality}
\end{align}
where the right-hand side should be interpreted as the zero matrix, i.e., every entry is zero. 

The frequency-domain $\TT$ matrix is the Fourier transform of the time-domain $\TT$ matrix:
\begin{align}
    \TT(\omega) &= \frac{1}{2\pi} \int_{-\infty}^{\infty} \TT(t) e^{i\omega t} \,{\rm d}t \nonumber \\
                &= \frac{1}{2\pi} \int_0^{\infty} \TT(t) e^{i\omega t} \,{\rm d}t,
                \label{eq:TTw}
\end{align}
where the integrand starts at zero in the second line because of the causality condition of \eqref{causality}. We assume the integral in \eqref{TTw} is well-behaved for all real frequencies. Then, if one inserts a complex-valued frequency into the expression of \eqref{TTw}, it is guaranteed to converge (under suitable conditions~\cite{Nussenzveig1972}), which leads to the usual property of complex-analyticity in the upper half of the complex-frequency plane, now for each entry of the $\TT$ matrix.

Once we have complex-analyticity, we can consider an integral of the form
\begin{align}
    \int_C \frac{\TT(\omega')}{\omega'- \omega} \,{\rm d}\omega',
    \label{eq:CI}
\end{align}
where $C$ is the typical semicircular contour in the complex plane with a small deviation around the frequency $\omega$. The integral in \eqref{CI} has a simple pole at $\omega$. Assuming sufficient decay at infinite frequencies in the upper-half plane (where physical materials become transparent and the $\TT$ matrix goes to zero), the semicircular arc in the upper-half plane does not contribute to the integral, and one is left only with the principal-valued integral along the real line at all points except $\omega$, and the residue term at $\omega$ given by:
\begin{align}
    \int_{\omega} \frac{\TT(\omega')}{\omega' - \omega} \,{\rm d}\omega' = -i\pi \TT(\omega).
\end{align}
The total integral in \eqref{CI} equals zero, so separating the principal-value and residue terms and equating their negatives, we have (assuming principal-value integrals where needed):
\begin{align}
    \int_{-\infty}^{\infty} \frac{\TT(\omega')}{\omega' - \omega} \,{\rm d}\omega' = i\pi \TT(\omega).
    \label{eq:Cint}
\end{align}
This is a matrix-valued equation that applies entrywise on each side. We can ``take the imaginary part of each side,'' really meaning that on each side we take the difference between the current term and its conjugate transpose, and divide by $2i$. This ``imaginary part'' of \eqref{Cint} then reads (moving the terms on the left to the right and vice versa),
\begin{align}
    \pi \Re \TT(\omega) = \int_{-\infty}^{\infty} \frac{\Im \TT(\omega')}{\omega' - \omega} \,{\rm d}\omega',
    \label{eq:Cint2}
\end{align}
where, as in the main text, ``$\Re$'' and ``$\Im$'' denote the Hermitian and anti-Hermitian parts of their matrix arguments. 

Finally, we can use symmetry of the $\TT$ operator around the origin. For an electric field or a polarization field, the usual symmetry relation is $\Ev(-\omega) = \Ev^*(\omega)$ and $\Pv(-\omega) = \Pv^*(\omega)$, which arises via the real-valued nature of the time-domain fields and the Fourier-transform definition of the frequency-domain fields. By exactly the same reasoning for the $\TT$ matrix, we have:
\begin{align}
    \TT_{ij}(-\omega) = \TT_{ij}^{*}(\omega).
    \label{eq:TTsym}
\end{align}
Because of reciprocity, $T_{ij} = T_{ji}$, we can equivalently write $\TT(-\omega) = \TT^\dagger(\omega)$, which implies a symmetry in the matrix imaginary part of $\TT$:
\begin{align}
    \Im \TT(-\omega) = -\Im \TT(\omega).
\end{align}
Hence the integral over the real line can be simplified to positive frequencies:
\begin{align}
    \int_{-\infty}^{\infty} \frac{\Im \TT(\omega')}{\omega' - \omega} \,{\rm d}\omega' &= \int_{-\infty}^{\infty} \frac{(\omega' + \omega) \Im \TT(\omega')}{(\omega')^2 - \omega^2} \,{\rm d}\omega' \nonumber \\
    &= \int_{-\infty}^{\infty} \frac{\omega' \Im \TT(\omega')}{(\omega')^2 - \omega^2} \,{\rm d}\omega' \nonumber \\
    &= 2\int_0^{\infty} \frac{\omega' \Im \TT(\omega')}{(\omega')^2 - \omega^2} \,{\rm d}\omega'.
\end{align}
Now we have our ``KK relation:''
\begin{align}
    \Re \TT(\omega) = \frac{2}{\pi} \int_0^{\infty} \frac{\omega' \Im \TT(\omega')}{(\omega')^2 - \omega^2} \,{\rm d}\omega'.
\end{align}

\section{$\TT$ matrix KK relation and representation theorem: Nonreciprocal Case}
In this section we identify the KK relation, and general representation theorem, for a scattering $\TT$ matrix in the case when reciprocity does \emph{not} apply.

First, we want a ``KK relation.'' In this case, we can just use \eqref{Cint2}:
\begin{align}
    \Re \TT(\omega) = \frac{1}{\pi} \int_{-\infty}^{\infty} \frac{\Im \TT(\omega')}{\omega' - \omega} \,{\rm d}\omega'.
\end{align}
This, already, is nearly an oscillator representation, including both positive and negative oscillator frequencies. We want the expression in the numerator to be positive semidefinite, but in this case, that is only true at positive frequencies. However, $\omega' \Im \TT(\omega')$ is positive
semidefinite across all frequencies, positive and negative. So we can multiply by $\omega'$ to find:
\begin{align}
    \Re \TT(\omega) = \frac{1}{\pi} \int_{-\infty}^{\infty} \frac{\omega' \Im \TT(\omega')}{\omega'(\omega' - \omega)} \,{\rm d}\omega'.
    \label{eq:TTallW}
\end{align}
This will form the basis for our oscillator representation, and can be regarded as a mathematical-oscillator representation in nonreciprocal (as well as reciprocal) systems.

Next we need a passivity condition and sum rules. For passive systems, the polarization fields do no net work, which implies that
\begin{align}
    \omega' \Im \TT(\omega') \geq 0,
\end{align}
at all (positive and negative) frequencies.

Lastly, we need sum rules. The zero-frequency sum rule can be derived directly from \eqref{TTallW}; if we denote $\Re \TT(\omega=0) = \alpha$, then
\begin{align}
    \int_{-\infty}^{\infty} \frac{\Im \TT(\omega')}{\omega'} \,{\rm d}\omega' = \pi \alpha.
\end{align}
A high-frequency sum rule is harder to identify. The intuition is as follows: we can decompose any $\TT$ matrix into a ``reciprocal'' component, that is complex-symmetric, and a ``nonreciprocal'' (or ``anti-reciprocal'' component), which is skew-symmetric, at every frequency. The reciprocal component satisfies the matrix analogs of the usual symmetry rules around the imaginary frequency axis, while the nonreciprocal component satisfies the opposite. Over all negative and positive frequencies, the reciprocal component will comprise the entire sum rule, while the nonreciprocal component will contribute zero. Let's work through this mathematically.

We start with the symmetry relation about the origin, \eqref{TTsym}. The key point about \eqref{TTsym} is that it is \emph{entrywise}, i.e., each entry satisfies its own symmetry relation. This prohibits direct symmetry relations for the Hermitian and anti-Hermitian parts of $\TT$ in the nonreciprocal case. (In the reciprocal case, we could use the transpose operator to turn the conjugation into a conjugate transpose operation.)

The key, then, is to break the $\TT$ matrix into its complex-symmetric and skew-symmetry parts, which we can refer to as its ``reciprocal'' and ``nonreciprocal'' (or really, ``anti-reciprocal'') parts:
\begin{align}
    \TT = \XX + \YY,
\end{align}
where
\begin{align}
    \XX &= \frac{\TT + \TT^T}{2},  \nonumber\\
    \YY &= \frac{\TT - \TT^T}{2}.
\end{align}
This is useful because now each component obeys \emph{matrix} symmetry relations:
\begin{align}
    \XX_{ij}(-\omega) = \frac{\TT_{ij}(-\omega) + \TT_{ji}(-\omega)}{2} = \frac{\TT_{ij}^*(\omega) + \TT_{ji}^*(\omega)}{2} = \XX_{ij}^*(\omega) = \XX_{ji}^*(\omega) \\
    \YY_{ij}(-\omega) = \frac{\TT_{ij}(-\omega) - \TT_{ji}(-\omega)}{2} = \frac{\TT_{ij}^*(\omega) - \TT_{ji}^*(\omega)}{2} = \YY_{ij}^*(\omega) = -\YY_{ji}^*(\omega),
\end{align}
which, in matrix form, can be written: 
\begin{align}
    \XX(-\omega) &= \XX^\dagger(\omega) \nonumber \\
    \YY(-\omega) &= -\YY^\dagger(\omega). \nonumber
\end{align}
From these,
\begin{align}
    \Re \XX(-\omega) &= \Re \XX(\omega) \nonumber \\
    \Im \XX(-\omega) &= - \Im \XX(\omega), \nonumber \\
    \Re \YY(-\omega) &= - \Re \YY(\omega) \nonumber \\
    \Im \YY(-\omega) &=  \Im \YY(\omega).
    \label{eq:XYsym}
\end{align}
If we look for a sum rule for the quantity
\begin{align}
    \int_{-\infty}^{\infty} \omega \Im \TT(\omega) \,{\rm d}\omega,
\end{align}
then we can insert the symmetric and skew-symmetric parts of $\TT$ into this expression, and use the symmetry relations of \eqref{XYsym} to find
\begin{align}
    \int_{-\infty}^{\infty} \omega \Im \TT(\omega) \,{\rm d}\omega = \int_{-\infty}^{\infty} \left[\omega \Im \XX(\omega) + \omega \Im \YY(\omega)  \right] \,{\rm d}\omega
\end{align}
The first integrand, $\omega \Im \XX(\omega)$, is symmetric around the origin, while the second is anti-symmetry and thus integrates to zero. Hence,
\begin{align}
    \int_{-\infty}^{\infty} \omega \Im \TT(\omega) \,{\rm d}\omega = 2\int_0^{\infty} \omega \Im \XX(\omega) \,{\rm d}\omega.
    \label{eq:XTrel}
\end{align}
Then we only need to find a sum rule for the symmetric (reciprocal) part of $\TT(\omega)$ to get a sum rule over all frequencies for $\omega \Im \TT(\omega)$.

To find a sum rule for $\XX(\omega)$, it is useful to find a \emph{separate} KK relation for this component. We can do this from the KK relation for $\TT(\omega)$, \eqref{Cint}. We can write this down for two entries of the $\TT$ matrix: the $ij$ entry and the $ji$ entry:
\begin{align}
    \TT_{ij}(\omega) &= \int_{-\infty}^{\infty} \frac{\TT_{ij}(\omega')}{\omega' - \omega} \,{\rm d}\omega'. \\
    \TT_{ji}(\omega) &= \int_{-\infty}^{\infty} \frac{\TT_{ji}(\omega')}{\omega' - \omega} \,{\rm d}\omega'.
\end{align}
We can add these two equations together to cancel the skew-symmetric parts, leaving only
\begin{align}
    \XX(\omega) &= \frac{1}{i\pi}\int_{-\infty}^{\infty} \frac{\XX(\omega')}{\omega' - \omega} \,{\rm d}\omega'.
\end{align}
This is a KK relation for $\XX(\omega)$; moreover, $\XX(\omega)$ obeys the anticipated \emph{matrix} symmetry relations around the imaginary axis, so we can write
\begin{align}
    \Re \XX(\omega) &= \frac{1}{\pi}\int_{-\infty}^{\infty} \frac{(\omega'+\omega)\Im \XX(\omega')}{(\omega')^2 - \omega^2} \,{\rm d}\omega' \\
                    &= \frac{1}{\pi}\int_{-\infty}^{\infty} \frac{\omega'\Im \XX(\omega')}{(\omega')^2 - \omega^2} \,{\rm d}\omega' \\
                    &= \frac{2}{\pi}\int_0^{\infty} \frac{\omega'\Im \XX(\omega')}{(\omega')^2 - \omega^2} \,{\rm d}\omega'.
                    \label{eq:KKX}
\end{align}
Next, we use the high-frequency asymptote of $\TT(\omega)$, $\TT(\omega) \rightarrow -\omega_p^2 / \omega^2 \II$ as $\omega \rightarrow \infty$ to write
\begin{align}
    \XX(\omega) \rightarrow -\frac{\omega_p^2}{\omega^2} \II \qquad \textrm{ as } \omega \rightarrow \infty.
\end{align}
Inserting this into the KK relation for $\XX$, \eqref{KKX}, we get, without any assumption of reciprocity:
\begin{align}
    \int_0^\infty \omega \Im \XX(\omega) \,{\rm d}\omega = \frac{\pi\omega_p^2}{2} \II.
\end{align}
We know this integral determines the integral of $\omega \Im \TT(\omega)$, and indeed can insert it into \eqref{XTrel} to find:
\begin{align}
    \int_{-\infty}^{\infty} \omega \Im \TT(\omega) \,{\rm d}\omega = \pi \omega_p^2 \II.
\end{align}
Now, over negative and positive frequencies, we have our three needed items: a positive semidefinite quantity, a Kramers--Kronig relation in terms of that positive-semidefinite quantity, and a sum rule (both zero- and high-frequency). We can compile here in one place:
\begin{align}
    \Re \TT(\omega) = \frac{1}{\pi} \int_{-\infty}^{\infty} \frac{\omega' \Im \TT(\omega')}{\omega'(\omega' - \omega)} \,{\rm d}\omega'. \\
    \int_{-\infty}^{\infty} \omega \Im \TT(\omega) \,{\rm d}\omega = \pi \omega_p^2 \II. \\
    \omega \Im \TT(\omega) \geq 0.
    \label{eq:TrepNR}
\end{align}

At first blush, it might seem from the sum rule, which now has $\pi\omega_p^2$ instead of $\pi\omega_p^2/2$ on its right-hand side, like you should be able to do a factor of two (or more) better in the nonreciprocal case than in the reciprocal case. But the symmetry condition on $\TT$, that each entry equal the conjugate of its negative-frequency counterpart, imposes a strict condition on $\omega \Im \TT(\omega)$:
\begin{align}
    -\omega \Im \TT(-\omega) = \left[\omega \Im \TT(\omega)\right]^T.
\end{align}
Hence, we must always put a concomitant oscillator at negative frequencies when an oscillator is placed at positive frequencies. Hence, for any optimization problem over all possible oscillator strengths, one should also impose this symmetry condition.

But for just about any objective, there is an even simpler optimization trick we can play to enforce this symmetry constraint. Instead of enforcing the constraint in the oscillator strength itself, we can enforce it in the objective. The fields $\Ev$, $\Hv$, $\Pv$, etc., all satisfy the usual symmetry condition $s(-\omega) = s^*(\omega)$, and any objective made of these functions appears to be symmetric about the origin. We can use Poynting flux as an example:
\begin{align}
    f(\omega) = \frac{1}{2} \Re \int \Ev(\omega) \times \Hv^*(\omega) \cdot \hat{\vect{n}}.
\end{align}
Its negative-frequency counterpart is
\begin{align}
    f(-\omega) &= \frac{1}{2} \Re \int \Ev(-\omega) \times \Hv^*(-\omega) \cdot \hat{\vect{n}}, \nonumber \\
               &= \frac{1}{2} \Re \int \Ev^*(\omega) \times \Hv(\omega) \cdot \hat{\vect{n}}, \nonumber \\
               &= \frac{1}{2} \Re \int \Ev(\omega) \times \Hv^*(\omega) \cdot \hat{\vect{n}} \nonumber \\
               &= f(\omega).
\end{align}
Because of this, instead of optimizing a positive-frequency objective subject to the constraint of symmetry in the oscillator strength, we can drop the constraint of symmetry in the oscillator strength, and instead maximize any linear combination of the negative- and positive-frequency objectives:
\begin{align}
    F(\omega) = \alpha f(\omega) + (1-\alpha) f(-\omega).
\end{align}
Typically it may be sufficient to simply choose $\alpha = 1/2$. Hence there are no additional degrees of freedom in the negative frequencies. 

One more version of a representation theorem in the nonreciprocal case can be specified, and this case most neatly allows comparison with the representation theorem in the reciprocal case. We showed above that there is a KK relation for $\XX(\omega)$, the complex-symmetric part of $\TT$, \eqref{KKX}. Similarly there is a KK relation for $\YY(\omega)$; it follows the same derivation, but now has the opposite symmetry around the origin, so that the final relation becomes
\begin{align}
    \Re \YY(\omega) = \frac{2}{\pi} \int_0^{\infty} \frac{\omega \YY(\omega')}{(\omega')^2 - \omega^2} \,{\rm d}\omega'.
    \label{eq:KKY}
\end{align}
Notice that the term in the numerator of the fraction in the integrand has $\omega$ now, and not $\omega'$. We can put the KK relations for $\XX$ and $\YY$ together to get a positive-frequency-only representation of $\TT(\omega)$:
\begin{align}
    \Re \TT(\omega) = \frac{2}{\pi} \int_0^{\infty} \frac{\omega' \Im \XX(\omega') + \frac{\omega}{\omega'} \omega' \Im \YY(\omega')}{(\omega')^2 - \omega^2} \,{\rm d}\omega',
\end{align}
where we have intentionally written the numerator of the integrand in terms of the component parts of $\omega' \Im \TT(\omega') = \omega' \Im \XX(\omega') + \omega' \Im \YY(\omega')$. Let us define two new matrices $\UU$ and $\VV$ such that
\begin{align}
    \UU(\omega) &= \omega \Im \XX(\omega) \\
    \VV(\omega) &= \omega \Im \YY(\omega),
\end{align}
which inherit the sum rules and symmetries of $\XX$ and $\YY$, such that $\UU$ is symmetric about the frequency origin while $\VV$ is anti-symmetric about the frequency origin. Moreover $\VV$ is real-symmetric (in space) while $\VV$ is anti-symmetric. We do not necessarily know that whether each individual matrix is positive semidefinite, but we know that their sum is, at all positive and negative frequencies. Hence, for a positive frequency $\omega$,
\begin{align}
    \UU(\omega) + \VV(\omega) \geq 0, \textrm{ and} \\
    \UU(-\omega) + \VV(-\omega) \geq 0.
\end{align}
The second condition, using the symmetry relations about the origin, can be converted to $\UU(\omega) - \VV(\omega) \geq 0$. Together, these two conditions imply that $\UU$ is positive semidefinite, $\UU(\omega) \geq 0$. So we now have:
\begin{align}
    \Re \TT(\omega) &= \frac{2}{\pi} \int_0^{\infty} \frac{\UU(\omega') + \frac{\omega}{\omega'} \VV(\omega') }{(\omega')^2 - \omega^2} \,{\rm d}\omega', \nonumber \\
    \UU(\omega) &\geq 0, \nonumber \\
    \UU(\omega) + \VV(\omega) &\geq 0, \nonumber \\
    \UU(\omega) - \VV(\omega) &\geq 0. \nonumber \\
    \int_0^{\infty} \UU(\omega) \,{\rm d}\omega &= \frac{\pi\omega_p^2}{2}\II.
\end{align}
For nonreciprocal problems, then, we get to choose two matrices at every frequency: one real-symmetry and one skew-symmetric. The real-symmetric matrix must be positive definite and has an all-positive-frequency sum rule. The skew-symmetry part has no sum rule but must satisfy two positivity conditions, in tandem with the symmetric part, at all frequencies. The skew-symmetric matrix is the only new degree of freedom in nonreciprocal systems.

\section{Low-frequency sum rule for the $\TT$ matrix}
\label{sec:sum_rule_TT0}
In the main text, we discussed the simple method for deriving a high-frequency sum rule for $\omega \Im \TT(\omega)$, by taking the pole frequency to infinity and using the transparency of materials at high frequencies to simplify the analysis. The other special frequency at which the analysis is simplified is zero, where the scattering problem becomes electrostatic. In this section, we derive the low-frequency sum rule, and the corresponding representation theorem.

As a reminder, the KK relation for $\TT(\omega)$ reads
\begin{align}
    \Re \TT(\omega) = \frac{2}{\pi} \int_0^{\infty} \frac{\omega' \Im \TT(\omega')}{(\omega')^2 - \omega^2} \,{\rm d}\omega'.
\end{align}
To derive a low-frequency sum rule, we simply evaluate this expression at $\omega = 0$:
\begin{align}
    \Re \TT(\omega=0) = \frac{2}{\pi} \int_0^{\infty} \frac{\Im \TT(\omega')}{\omega'} \,{\rm d}\omega'.
\end{align}
The matrix $\TT(\omega=0)$ at zero frequency is  a generalized polarizability matrix (since it relates the induced dipole density to the incident field), which we can denote by $\alphat$. This matrix is real-symmetric per the discussion in the previous section: decomposing $\TT(\omega)$ into its symmetric ($\XX$) and anti-symmetric ($\YY$) parts, the latter has zero Hermitian part at zero frequency. Now our sum rule reads
\begin{align}
    \int_0^{\infty} \frac{\Im \TT(\omega')}{\omega'} \,{\rm d}\omega' = \frac{\pi}{2} \alphat,
    \label{eq:SRTT0}
\end{align}
which further shows that the generalized polarizability tensor is positive-semidefinite (since the left-hand side is).

As discussed in \secref{monotonicity}, this generalized polarizability tensor obeys ``domain monotonicity,'' meaning that the difference between the polarizability tensor of a domain and any second domain that is enclosed by the first must be positive-definite. Because of this, we can typically choose a high-symmetry enclosure (e.g. half-spaces) for which the polarizability is a scalar $\alpha$ multiplied by the identity matrix, and \emph{all possible geometric domains inside of them will have sums that satisfy}:
\begin{align}
    \int_0^{\infty} \frac{\Im \TT(\omega')}{\omega'} \,{\rm d}\omega' \leq \frac{\pi\alpha}{2} \II.
    \label{eq:SRTT0b}
\end{align}

From the sum rule of \eqref{SRTT0b}, we can form a representation of $\TT$ similar to the high-frequency representation of the main text. If we start by choosing $\Im \TT(\omega) / \omega$ to be a summation of delta functions, then we find a natural representation of $\omega \Im \TT(\omega)$ as
\begin{align}
    \omega \Im \TT(\omega) = \sum_i \frac{\pi\alpha\omega_i^2}{2} \delta(\omega - \omega_i),
\end{align}
where we used the fact that $\omega^2 \delta(\omega - \omega_i) = \omega_i^2 \delta(\omega-\omega_i)$. Now we can see that this oscillator representation is exactly the same as in the high-frequency case, except with the replacement $\omega_p^2 \rightarrow \alpha \omega_i^2$. Hence the representation theorem will be the same, but with the coefficients replaced:
\begin{align}
    \TT(\omega) = \lim_{\gamma \rightarrow 0} \sum_i \frac{\alpha \omega_i^2}{\omega_i^2 - \omega^2 - i\gamma\omega} \TT_i,
    \label{eq:Trep0}
\end{align}

In the NFRHT bound, we used the generalized polarizability for two half-spaces, $\alpha_{\rm 2hs} = 2$. To derive this result, we first consider a simple case of a single half-space interface parallel to the $xy$-plane, and the medium in $z<0$ has permittivity $\varepsilon_1 = 1$ while the medium in $z>0$ is our half-space scatterer that has permittivity $\varepsilon_2$.
A general electrostatic source is located in the air side, and at $z = -d$ away from the interface.
In electrostatics, away from the source and the interface, one can write ${\bm E} = - \nabla \psi$ where $\nabla^2\psi = 0$.
At each $z$, $\psi$ can be expressed with a 2D Fourier integral:
\begin{align}
    \psi(x,y,z) = \iint \limits_{-\infty}^{+\infty} {\rm d} k_x {\rm d} k_x \; \tilde{\psi} \left(k_x, k_y, k_z \right) e^{i k_x x + i k_y y},
\end{align}
where $\tilde{\psi}$ is the 2D Fourier transform of $\psi$. Away from the source and the interface, one can solve the electrostatic Poisson's equation and obtain the expressions for the electric field:
\begin{align}
    {\bm E}(x,y,z) &= \iint \limits_{-\infty}^{+\infty} {\rm d} k_x {\rm d} k_x \; \left(k_x,k_y,k_z \right) U\left(k_x,k_y \right)e^{i k_x x + i k_y y + i k_z z}\\
    &+  \iint \limits_{-\infty}^{+\infty} {\rm d} k_x {\rm d} k_x \; \left(k_x,k_y,-k_z \right) V\left(k_x,k_y \right)e^{i k_x x + i k_y y - i k_z z}
\end{align}
for $-d < z < 0$, and 
\begin{align}
    {\bm E}(x,y,z) &= \iint \limits_{-\infty}^{+\infty} {\rm d} k_x {\rm d} k_x \; \left(k_x,k_y,k_z \right) W\left(k_x,k_y \right)e^{i k_x x + i k_y y + i k_z z}
\end{align}
for $z>0$, where $U$, $V$, and $W$ are the plane-wave modal field amplitudes for the incoming, the reflected and the transmitted fields.
Note that in electrostatics, not only $k_x$ and $k_y$ but also $k_z = i\sqrt{k^2_x+k^2_y}$ are conserved across the interface.

To find $\TT(\omega = 0)$, we need to find the relation between the polarization current ${\bm P} = \chi {\bm E}$ and the incident field ${\bm E_{\rm inc}}$ in the region $z>0$, which is essentially finding the Fresnel coefficients. What are the Fresnel coefficients in electrostatics? As pointed out in ~\citeasnoun{static_Fresnel}, Fresnel equations apply to statics, and for electrostatic sources:
\begin{align}
    r &= \frac{V}{U} = \frac{\varepsilon_1-\varepsilon_2}{\varepsilon_1+\varepsilon_2}\\
    t &= \frac{W}{U} = \frac{2\varepsilon_1}{\varepsilon_1+\varepsilon_2}
\end{align}
Importantly, note that the Fresnel coefficients are independent of $k_x$ and $k_y$, and therefore after the inverse Fourier transform, we have $\bm E = t{\bm E_{\rm inc}}$, and $\alpha= \chi t$. Similarly, the arguments expressing the fields with Fourier basis apply when we consider two parallel half-spaces separated by $d_0$, but the transmission coefficient needs to be substituted by that of two interfaces
\begin{align}
    t_{\rm 2hs} &= \frac{t\left( 1+ r e^{2 i k_z d} \right)}{1 - r^2 e^{2 i k_z d_0}}
\end{align}
Using $\varepsilon_2 = 1 - \frac{\omega^2_p}{\omega^2}$ at $\omega \rightarrow 0$, one can obtain $t_{\rm 2hs}(\omega = 0)=\frac{2}{\varepsilon_2}$ and $\alpha_{\rm 2hs} = \chi t_{\rm 2hs} = 2$. Therefore the electrostatic $\TT$ matrix for the bounding volume of two half-spaces is $\TT(\omega = 0) = \alpha_{\rm 2hs} \II$ where $\alpha_{\rm 2hs} = 2$.

\section{Drude--Lorentz representation of the $\TT$ matrix}
Here we discuss some of the mathematical aspects of our Drude--Lorentz oscillator representation in the main text. We started with three identities,
\begin{align}
    &\Re \TT(\omega) = \frac{2}{\pi} \int_0^{\infty} \frac{\omega' \Im \TT(\omega')}{(\omega')^2 - \omega^2} \,{\rm d}\omega'. \\
    \int_0^{\infty} &\omega' \Im \TT(\omega') \,{\rm d}\omega' = \frac{\pi\omega_p^2}{2}. \\
                    &\omega \Im \TT(\omega) \geq 0 \qquad \textrm{ for all } \omega.
\end{align}
From these three identities, we represented the imaginary parts of the scattering matrix via delta functions at a discrete set of points along the real line, from 0 to arbitrarily high frequencies:
\begin{align}
    \omega \Im \TT(\omega) = \frac{\pi \omega_p^2}{2} \sum_i \TT_i \delta(\omega - \omega_i).
    \label{eq:Tdisc}
\end{align}
Then we claimed that this leads to a single unified representation,
\begin{align}
    \TT(\omega) = \lim_{\gamma \rightarrow 0} \sum_i \frac{\omega_p^2}{\omega_i^2 - \omega^2 - i\gamma\omega} \TT_i,
    \label{eq:KKT}
\end{align}
where the $\TT_i$ satisfy $\TT_i \geq 0$ and $\sum_i \TT_i \leq \II$. First, we can simply validate that this expression is indeed consistent with the three identities above. The positivity and sum-rule conditions are clearly met. We next want to ensure that the real and imaginary parts are correct. In the discrete representation of \eqref{Tdisc}, the resulting real part is given by 
\begin{align}
    \Re \TT(\omega) = \sum_i \frac{\omega_p^2 }{\omega_i^2 - \omega^2} \TT_i.
\end{align}
Clearly this is consistent with \eqref{KKT} in the limit as $\gamma \rightarrow 0$. (We will see another validation in a moment.) Is the imaginary part of \eqref{KKT} consistent with the delta function representation? One way to see this would be to separate the real and imaginary parts of \eqref{KKT}, see that the linewidth of the imaginary part is going to zero but that its integral is fixed at 1. An alternative route is to use partial fractions to rewrite
\begin{align}
    \lim_{\gamma \rightarrow 0} \frac{1}{\omega_i^2 - \omega^2 - i\gamma\omega} &= \lim_{\gamma \rightarrow 0}\left\{ \frac{1}{2\omega_i} \left[ \frac{1}{\omega_i - \omega - i\gamma/2} - \frac{1}{-\omega_i - \omega - i\gamma/2} \right]\right\} \\
                                                                                &= \frac{1}{2\omega_i} \left[ \operatorname{PV}\left(\frac{1}{\omega+\omega_i}\right) + \operatorname{PV}\left(\frac{1}{\omega_i - \omega}\right) + i\pi\delta(\omega_i - \omega) - i\pi\delta(-\omega_i-\omega) \right],
                                                                                \label{eq:PVrep}
\end{align}
where ``PV'' denotes principal value, and to reach the second line we used the well-known Sokhotski--Plemelj identity~\cite{Nussenzveig1972}
\begin{align}
    \lim_{\varepsilon\rightarrow 0^{+}} \left( \frac{1}{\omega' - \omega - i\varepsilon} \right) = \operatorname{PV}\left(\frac{1}{\omega' - \omega}\right) + i\pi \delta(\omega' - \omega).
\end{align}
We can insert the expression of \eqref{PVrep} into the expression for $\TT(\omega)$, \eqref{KKT}, to find:
\begin{align}
    \TT(\omega) = \omega_p^2 \sum_i \left\{ \operatorname{PV}\left(\frac{1}{\omega_i^2 - \omega^2}\right) + \frac{i\pi}{2\omega} \left[ \delta(\omega_i - \omega) + \delta(\omega_i+\omega) \right] \right\} \TT_i,
\end{align}
where to simplify the last term we used the identities $\delta(\omega_i - \omega)/\omega_i = \delta(\omega_i - \omega)/\omega$ and $\delta(-x) = \delta(x)$. Clearly we can see that
\begin{align}
    \Re \TT(\omega) &= \sum_i \operatorname{PV}\left(\frac{\omega_p^2}{\omega_i^2 - \omega^2} \right) \TT_i, \textrm{ and} \\
    \omega \Im \TT(\omega) &= \sum_i \frac{\pi\omega_p^2}{2}\TT_i\delta(\omega - \omega_i),
\end{align}
exactly as required. Hence we have shown that the oscillator representation of \eqref{KKT} encodes all of the identities we derived.

For simplicity of exposition and intuition we used the basis of delta functions. One can expect arbitrarily high accuracy using this approach when working with objectives that satisfy two conditions: first, integration over bandwidths at least as large as the (infinitesimal) oscillator strengths $\gamma$, and second, an oscillator spacing well below any other characteristic oscillations (in the fields, target objectives, etc.) in the system. In the few cases where such conditions cannot be met in a computationally expedient calculation, there are two other routes that one can take. First, we can borrow from techniques used for material Kramers--Kronig-based optimizations. Starting from the Kramers--Kronig relations, instead of discretizing in delta functions, one can instead discretize in a basis of spline functions~\cite{Ivanenko2019}. Spline functions have analytical Hilbert transforms, which make them a very convenient basis to use.

A second alternative approach is to use the theory of Herglotz (or Nevanlinna, or Pick, or Herglotz--Nevanlinna-Pick) functions~\cite{Nussenzveig1972,King2009,Bernland2011,Luger2022,Ou2022}. A scalar function $f(z)$ is a ``Herglotz function'' if it satisfies two simple properties in the upper-half place (UHP):
\begin{align}
    &f(z) \textrm{ is holomorphic in the UHP, and} \nonumber \\
    \Im &f(z) \geq 0 \textrm{ in the UHP.} 
    \label{eq:Herglotz}
\end{align}
The foundational relation for a Herglotz function is that the two properties of \eqref{Herglotz} are sufficient to require a representation of $f(z)$ of the following form~\cite{Nussenzveig1972,King2009,Luger2022}:
\begin{align}
    f(z) = Az + C + \int_{-\infty}^{\infty} \frac{1+tz}{t - z} \,{\rm d}\beta(t)
    \label{eq:Hsca}
\end{align}
where $z$ is in the UHP but traditionally taken in the limit of approaching the real line. The constants $A$ and $C$ are related to asymptotic behavior of $f(z)$, while ${\rm d}\beta(t)$ is a positive measure. This positive measure plays the role of $\omega \Im \chi$ in a discrete susceptibility oscillator representation, and can be transformed to precisely these quantities for which there are sum rules. Hence the fraction in the integrand plays the role of the ``oscillator,'' while the positive measure plays the role of the positive, sum-rule-constrained oscillator strength. A detailed derivation of the transformation of the Herglotz-function representation to the Kramers--Kronig relations for optical susceptibilities is given in \citeasnoun{King2006}.

The matrix-valued Herglotz representation natural extends the scalar case. Consider a matrix that is analytic in the UHP and whose imaginary (anti-Hermitian) part is positive-definite in the UHP. The $\TT$ matrix multiplied by frequency, i.e. $\omega \TT(\omega)$, satisfies these two conditions, as we showed in the main text. Then we are guaranteed to have a representation~\cite{Gesztesy2000,Fritzsche2012}
\begin{align}
    \omega \TT(\omega) = Az + C + \int_{-\infty}^{\infty} \frac{1+tz}{t - z} \nu({\rm d}t),
    \label{eq:Hmat}
\end{align}
where $A$ and $C$ are constant matrices and, crucially, $\nu$ is a positive semidefinite Hermitian measure. (In both the scalar- and matrix-valued cases, the positive measures satisfy bounded-integrability conditions to ensure that the integrals of \eqreftwo{Hsca}{Hmat} are well-behaved.) \Eqref{Hmat} is the continuous generalization of our discrete-summation representation of the $\TT$ matrix. The matrix $\nu$ plays the role of the positive-semidefinite oscillator strength, and indeed can be connected to the imaginary (anti-Hermitian) part of $\omega \TT(\omega)$ evaluated in the UHP, asymptotically approaching the real line~\cite{Gesztesy2000}. Then matrix-valued generalizations of the numerical techniques used for scalar Herglotz functions~\cite{Ivanenko2019} should be readily applicable.

\section{Scattering simulation parameters for Fig. 1}
In this section we provide the detailed simulation data and techniques for Fig. 1 of the main text. The elliptical cylinder has susceptibility $\chi = 4$, width $D_x = 2.4a$, and height $D_y = 1.6a$, where $a$ is a scale factor for length normalization. To obtain accurate results using the simulation method we will introduce below, the sharp edge of the geometry need to be smoothed. In this example, the susceptibility distribution of the elliptical cylinder is expressed as
\begin{align}
    \chi(x,y) = \frac{\chi_{\rm ell}}{2} \left\{ 1+ {\rm tanh} \left[ c_1 \left( 1 - \sqrt{\frac{x^2}{D_x^2}+\frac{y^2}{D_y^2}}  \right) \right]\right\},
    \label{eq:ellipse}
\end{align}
where $c_1$ is inversely proportional to the width of the smoothed area along the circumference of the ellipse.
For the full-wave simulation, we use our own direct solver utilizing a discrete dipole approximation (DDA) augmented by a Duan-Rokhlin quadrature~\cite{Yurkin2011DDA,Duan2009}. The simulation region is a square of side length $3.0a$. 
Discretization of the square region gives 512 grid points along both x and y direction. There are 501 frequency sampling points ranging from 0.02 to 1, in units of $2\pi c/a$. The $\TT$ matrix is obtained from \eqref{TTGF}, which is $\TT = -(\GG_0 + \xi\II)^{-1}$, where $\GG_0$ is the vacuum Green's function matrix and $\xi = -\frac{1}{\chi}$, both defined on the volume of scatterer. We use 6th-order Duan-Rokhlin correction for accurate computation of $G$, guaranteeing accuracy of less than 0.01\% error in the computed extinguished power of the structure, at all frequencies of interest.

For plotting the $\Escat$ and $\TT$ matrix elements, we select 5 random points inside the scatterer: $x_1 = (-0.79,-0.36)a$, $x_2 = (0.74,-0.12)a$, $x_3 = (0.17,-0.02)a$, $x_4 = (-0.40,0.07)a$, $x_5 = (0.93,0.31)a$, using the center of the ellipse as the origin. The incident field is a plane wave propagating along the $y$ direction with the electric field polarized perpendicular to the plane.

\section{Domain monotonicity: Why sharp tips do not enhance LDOS or NFRHT} \label{sec:monotonicity}
In this section, we explain why sharp tips, or any other geometric patterning, do not enhance the NFRHT bound.
As discussed in the main text, for any scattering body, the $only$ degrees of freedom are the scattering oscillator strengths $\TT_i$. The oscillator strengths are positive semidefinite, and their key constraint is the sum rule of \eqref{SRTT0}, which as discussed in \secref{sum_rule_TT0} is given by $\mathbb{T}(\omega = 0)$, the electrostatic $\TT$ matrix, which we also call the ``generalized polarizibility matrix'' ${\bm \alpha}$. The question, then, is what structure leads to the largest generalized polarizability / electrostatic $\TT$ matrix? In this section, we derive a ``domain montonicity theorem" for the electrostatic $\TT$ matrix, which shows that enclosing any domain with another will always ``increase" (in a matrix-valued sense made precise below) the electrostatic $\TT$ matrix. Then, among all structures with a minimum separation $d$, the largest electrostatic $\TT$ matrix will be that of the domain of two planar half-spaces. We prove the mathematics of this theorem below, and then discuss the physics: sharp tips enhance fields infinitely close to the tips themselves, but $not$ back at the location of the source itself. The latter is the key quantity for local densities of states and related quantities such as NFRHT.

To prove domain monotonicity, we need to prove that quantities of the form $\xv^\dagger \TT \xv$ increase, for all $\xv \neq 0$, when the domain of the $\TT$ matrix increases. We can interpret the multiplication of $\TT$ with $\xv$ as the polarization field induced by an ``incident field'' $\xv$, and then multiplication on the left by $\xv$ takes the overlap of that incident field with the polarization that it induces. Hence we will label our arbitrary vectors as $\einc$ instead of $\xv$, for clarity in the mathematical relations to follow, though we impose no constraints on the ``incident field'' and indeed allow it to be an arbitrary vector. In computing the \emph{response} to such a vector, however, we can use a few important physical consequences of electromagnetism. We will assume a scalar (and therefore reciprocal) electric susceptibility; the generalization to the arbitrary tensor case follows the same process as below in tandem with the modifications described in Sec. VI of the SM of \citeasnoun{Shim2019bandwidth}. In electrostatics the fields (and $\TT$ matrix) can be chosen to be real-valued, so that we can consider the objective as $\xv^T \TT \xv$, without any conjugation.

When an incident field $\einc$ interacts with a scatterer, for example through its $\TT$ scattering matrix, there is a scattered field produced. The volume-equivalence principle allows us to write the scattered field in two forms. The first is as the convolution of the induced polarization field $\pv$ with the background (vacuum) Green's function operator $\GG_0$, i.e. $\GG_0 \pv$. The second is a convolution of the full Green's function operator, $\GG$, with an overlap of the susceptibility $\chi$ with the incident $\einc$: $\escat = \GG \chi \einc$. The total field is the incident field plus the scattered field,
\begin{align}
    \ev = (\II + \GG \chi) \einc.
    \label{eq:IIGG}
\end{align}

We are interested in the quantity $F = \einc^T \TT \einc = \einc^T \pv$, and how it changes when the domain changes. We will consider only continuous, increasing changes in susceptibility: $\Delta \chi(\xv) \geq 0$ everywhere. Hence a variation in $F$ can be written
\begin{align}
    \delta F = \einc^T \delta \pv.
\end{align}
The change in the polarization field $\pv = \chi \ev$ across the scatterers has two contributions: a change in the permittity multiplied by the field, and the permittivity multiplied by the change in field:
\begin{align}
    \delta \pv = \left(\delta \chi\right) \ev + \chi (\delta \ev).
\end{align}
The change in field equals the convolution of the Green's function with the newly induced current, which is $(\delta \chi) \ev$. Hence we have
\begin{align}
    \delta \pv = \left(\delta \chi\right) \ev + \chi \GG (\delta \chi) \ev,
\end{align}
and,
\begin{align}
    \delta F &= \einc^T \left[ \left(\delta \chi\right) \ev + \chi \GG (\delta \chi) \ev \right] \nonumber \\
             &= \einc^T \left(\II + \chi \GG\right) \delta\chi \ev \nonumber \\
             &= \ev^T \delta\chi \ev \nonumber \\
             &\geq 0,
\end{align}
where to proceed from the second to the third line we used the previous scattering-matrix equality, \eqref{IIGG}, in tandem with the assumption of scalar (reciprocal) materials. The fourth line follows from the third because the fields are real and the change in susceptibility is positive, so 
\begin{align}
    \ev^T \delta \chi \ev = \int_V \delta\chi(\xv) |\Ev(\xv)|^2 \geq 0.
\end{align}
Hence we have shown that 
\begin{align}
    \einc^\dagger \delta \TT \einc \geq 0
\end{align}
for any increases in the domain size or shape; since this is true for any vector $\einc$, then variations in the electrostatic $\TT$ matrix must themselves be monotonic. This means that given a scatterer $\Omega_1$ of any size and shape whose static $\TT$ matrix is $\mathbb{T}^{(1)}(\omega = 0)$, any other scatterer $\Omega_2$ whose volume encloses that of $\Omega_1$ must have a $\mathbb{T}^{(2)}(\omega = 0)$ greater than $\mathbb{T}^{(1)}(\omega = 0)$, i.e.:
\begin{align}
\mathbb{T}^{(2)}(\omega = 0) > \mathbb{T}^{(1)}(\omega = 0),
\end{align}
when the scatterer domain $\Omega_2$ entirely encloses the scatter domain $\Omega_1$. Recalling from \eqref{SRTT0} that the electrostatic $\TT$ matrix determines the constant in the low-frequency sum rule for the corresponding structure, the domain monotonicity theorem dictates that the NFRHT bound obtains a similar monotonicity: the bound only increases for structures with larger bounding volumes. For any given gap sizes of NFRHT, two planar half-spaces provide the largest $\mathbb{T}(\omega = 0)$, which we use as the sum rule constant to derive the upper bound for NFRHT. Therefore, our bound applies to any structure, regardless of their size and shape, as they must satisfy the gap separation and therefore be contained by the two half-spaces. Nanopatterned sharp tips, despite potentially providing local field enhancement near the tip, must have smaller upper bounds due to smaller bounding volumes, according to the monotonicity theorem. 

To further illustrate why nano-patterning such as sharp-tip structure is inferior, we design a numerical experiment. The key objective function in the above analytical proof is the quantity $\einc^T \pv$, i.e. the overlap of an incident field with the polarization field it induces. In the case of NFRHT, the incident field arises from point sources along a separating plane between the two bodies. An incident field emanating from a dipolar point source $\pv_0$ can be written $\einc = \GG_0 \pv_0$, in which case the overlap can be written $\einc^T \pv = \pv_0^T \GG_0^T \pv = \pv_0^T \GG_0 \pv = \pv_0^T \escat$, where $\escat$ is the scattered field, emanating from the polarization fields, \emph{back} at the location of the source on the separating plane. Moreover, the overlap measures the zero-frequency response proportional to the component of ${\bm E_{\rm scat}}({\bm x_0})$ parallel to the polarization of the dipole source ${\bm p}({\bm x_0})$. We should pinpoint the static ${\bm E_{\rm scat}}({\bm x_0})$ as its magnitude is proportional to the electrostatic $\TT$ matrix.

We simulated the static ${\bm E_{\rm scat}}({\bm x_0})$ for a selection of representative structures to verify the monotonicity theorem. The simulations below uses the finite element method (FEM) of a commercial software, COMSOL Multiphysics. Point dipoles with unit amplitude are placed at the center of gaps of size $2d$ between identical wedges or half-spaces. The maximum size of triangular meshes are $0.01d$ in the central region of the air gap and around the tip, and $0.1d$ for the rest of region away from the dipole. ${\bm E_{\rm scat}}({\bm x_0})$ is computed for vertical ``$\bot$'' (perpendicular to the separating plane) and horizontal ``$\parallel$'' (parallel to the separating plane) polarized ${\bm p}({\bm x_0})$ for 8 different structures with inner angles from $0.125\pi$ to $\pi$, essentially from a pair of sharp and pointed wedges to two planar half-spaces. Due to the intrinsic difficulty in measuring ${\bm E_{\rm scat}}({\bm x_0})$ which is at the point of dipole source location, it is instead obtained by interpolation of ${\bm E_{\rm scat}}({\bm x})$ at locations near the dipole source. The numerical data points for $\beta = \pi$ (two half-spaces) agree well with analytical prediction through method of images which gives $\Escat^{\bot}({\bm x_0}) = \frac{\pi}{24\epsn d^2}$ for vertical polarization and $\Escat^{\parallel}({\bm x_0}) = \frac{\pi}{48\epsn d^2}$ for horizontal polarization.

\begin{figure*}[htb]
  \includegraphics[width=\textwidth]{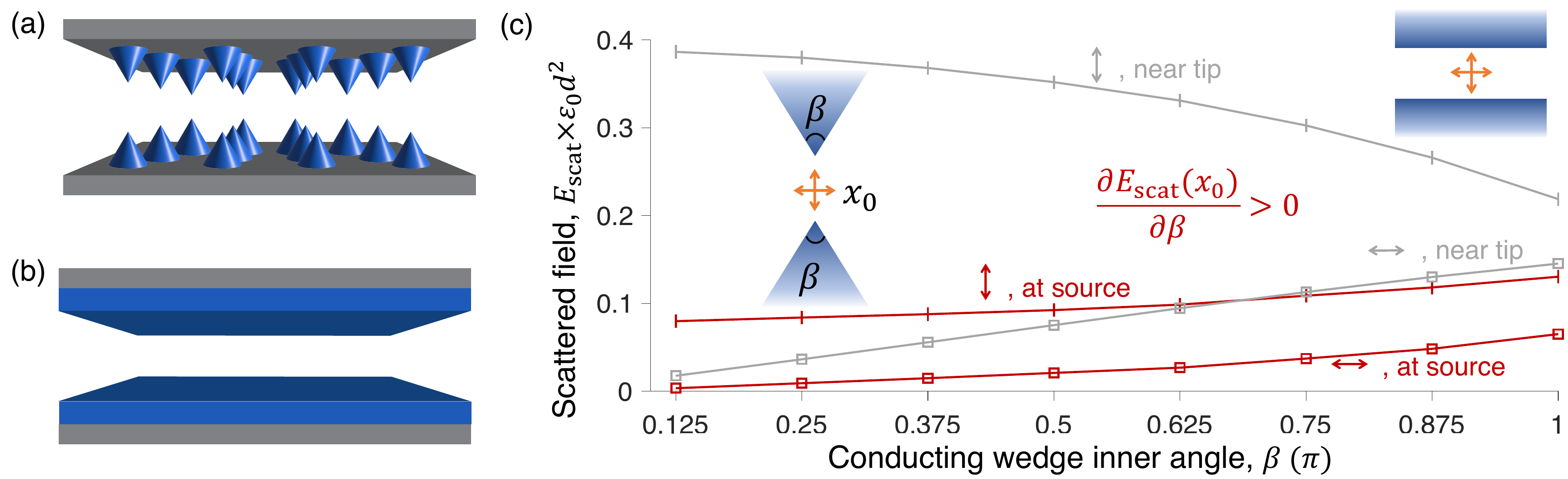}
  \caption{Domain monotonicity of static ${\bm E_{\rm scat}}({\bm x_0})$. (a) Structures patterned with sharp cones v.s. (b) planar structures without any patterning. (c) Scattered fields from a dipole placed in the middle of a pair of conducting wedges of different inner angles from $0.125\pi$ to $\pi$. Square markers signify horizontal dipole orientation, and vertical markers signify vertical dipole polarization. Despite there is monotonic increase of scattered field near the tip of the wedges with decreasing inner angles for the vertical polarization according to the top grey line, the two red lines, directly proportional to static ${\bm E_{\rm scat}}({\bm x_0})$, show monotonic increase of scattered field back at the source location with increasing inner angles. Largest static ${\bm E_{\rm scat}}({\bm x_0})$ is when $\beta = \pi$, namely the structure is two planar half-spaces.}
  \label{fig:monotonicity}
\end{figure*}

The results are shown in the figure \figref{monotonicity}(c) as dark red lines with vertical markers representing vertical polarization and square markers representing horizontal polarization.
Additionally, scattered field at a point close to the tip ($0.1d$ away from the tip along the line connecting two tips) are shown as grey lines. As wedge angle increases, the vertically polarized scattered field decreases near the tip and increases at source location, the horizontally polarized scattered field increases both near the tip and at source location.
In other words, although there is field enhancement near the tip with small wedge structures for the vertical polarization, which is the reasoning behind nano-patterning to improve near-field interaction and increase NFRHT efficiency, the quantity that is much more relevant, the scattered field right at the source point ${\bm E_{\rm scat}}({\bm x_0})$, is decreased with nano-patterning. The largest ${\bm E_{\rm scat}}({\bm x_0})$ is achieved with the largest wedge angle $\beta = \pi$, namely two unpatterned half-spaces. As the increase of wedge angle essentially increases scatterer domain size, this numerical result provides quantitative confirmation that static ${\bm E_{\rm scat}}({\bm x_0})$ is increasing as domain size monotonically increases. 

\section{Derivation of the NFRHT bound}
To investigate radiative heat transfer from object 1 (bottom) to object 2 (top), we first break down the problem to power integrations at every frequency. The power flowing in the positive $z$ direction across the middle separating plane (perpendicular to $z$) between the two objects is:
\begin{equation}
S_z(\omega,T) = \frac{1}{2} {\rm{Re}}\int {\rm{d}} S \left[ \left( E^{\bm{J}}_x(\bm{r_s})\right)^{*}H^{\bm{J}}_y(\bm{r_s}) - \left( E^{\bm{J}}_y(\bm{r_s})\right)^{*}H^{\bm{J}}_x(\bm{r_s}) \right]      , \\ 
    \label{eq:Sz}
\end{equation}	
where the superscripts denote the polarizations of the current sources in the bottom object, whose amplitudes are dictated by the fluctuation-dissipation theorem:
\begin{align}
\left< {J^{*}_i}\left( \omega,\bm{r_v}\right) {J_j}\left( \omega^{\prime},\bm{r^{\prime}_v}   \right)\right>_V &= \frac{4\en\omega}{\pi}  {\rm{Im}} \chi_1(\omega,\bm{r_v}) \Theta(\omega,T)  \delta(\omega-\omega^{\prime})   \delta(\bm{r_v}-\bm{r}^{\prime}_v)  \delta_{ij},
    \label{eq:JJ_correlation}
\end{align}
where the susceptibility of object 1 is $\chi_1(\omega) = \frac{\varepsilon_1(\omega)}{\en} -1$, and $\Theta(\omega,T)$ is the Planck distribution, $\Theta(\omega,T) = \frac{\hbar\omega} {e^{\frac{\hbar\omega}{\kb T}}-1 }$. The subscripts on the position vectors $\bm{r_s}$ and $\bm{r_v}$ indicate whether their domain is the surface (separating flux plane) or volume (of the emitter/absorber). Then the field correlations are found by rewriting the fields in terms of the  combining Green's function and thermal source correlations
\begin{align}
\left< \left(E^{J_i}_x(\bm{r_s})\right)^{*} H^{J_j}_y (\bm{r_s})\right>_V &= \left< \int {\rm d} v \left(i\omega\mu_0 G^{EJ}_{ik}(\bm{r_s},\bm{r_v})\right)^{*}\int {\rm d} v^{\prime} G^{EJ}_{ik}(\bm{r_s},\bm{r^{\prime}_v}) {J^{*}_i}\left( \omega,\bm{r_v}\right) {J_j}\left( \omega^{\prime},\bm{r^{\prime}_v}   \right)\right>.
    \label{eq:EH_correlation}
\end{align}

Our bound will not distinguish between the $x$ and $y$ directions (which are symmetric in the bounding domain, even though they of course are not for many allowable patterns), in which case the upper bounds on either of the two terms in power integration in \eqref{Sz} are identical:
\begin{equation}
{\rm{Max}} \left[ {\rm{Re}}\int {\rm{d}} S  \left( E^{\bm{J}}_x(\bm{r_s})\right)^{*}H^{\bm{J}}_y (\bm{r_s})\right] = {\rm{Max}} \left[- {\rm{Re}}\int {\rm{d}} S  \left( E^{\bm{J}}_y(\bm{r_s})\right)^{*}H^{\bm{J}}_x (\bm{r_s})\right]      . \\ 
\end{equation}	
Hence the maximum flux $S_z(\omega,T)$ equals the maximum of the function
\begin{equation}
F^{\rm{RHT}}(\omega,T) \equiv {\rm{Re}}\int {\rm{d}} S  \left( E^{\bm{J}}_x(\bm{r_s})\right)^{*}H^{\bm{J}}_y(\bm{r_s}).
    \label{eq:F_define}
\end{equation}
By Lorentz reciprocity,
\begin{equation}
G^{EJ}_{ik}(\bm{r},\bm{r^{\prime}}) = G^{EJ}_{ki}(\bm{r}^{\prime},\bm{r}),
\end{equation}
\begin{equation}
G^{HJ}_{ik}(\bm{r},\bm{r^{\prime}}) = - G^{EM}_{ki}(\bm{r}^{\prime},\bm{r}),
\end{equation}
we can equate the fields at $\bm{r_s}$ produced by sources at $\bm{r_v}$ with fields at $\bm{r_v}$ produced by sources at $\bm{r_s}$. This will be helpful to our bound formalism which would essentially be a volume integration of field operators. In light of the correlations for currents sources inside the volume, \eqref{JJ_correlation}, we can define the correlations for current sources on the middle flux plane as
\begin{equation}
\left<  {J^{*}_x}\left( \omega,\bm{r_s}\right) {M_y}\left( \omega^{\prime},\bm{r^{\prime}_s}\right)\right>_S  \equiv \omega \delta(\omega-\omega^{\prime})   \delta(\bm{r_s}-\bm{r_s}^{\prime}). \\ 
    \label{eq:JxMy_correlation}
\end{equation}
The amplitude $\omega$ is chosen so that $\left(E^{J_x}_{\rm inc}(\bm{r_v})\right)^{*} E^{M_y}_{\rm inc}(\bm{r_v})$ is independent of frequency, which will be important later. Now we can rewrite \eqref{EH_correlation} as
\begin{equation}
\left< \left(E^{J_i}_x(\bm{r_s})\right)^{*} H^{J_j}_y (\bm{r_s})\right>_V= \frac{4\en\omega}{\pi}  {\rm{Im}} \chi(\omega) \Theta(\omega,T) \frac{1}{\omega}  \delta_{ij}  \left<  \left(E^{J_x}_i(\bm{r_v})\right)^{*} E^{M_y}_j(\bm{r_v})\right>_S , \\ 
    \label{eq:Efield_correlation}
\end{equation}
and \eqref{F_define} as
\begin{equation}
F^{\rm{RHT}}(\omega,T) = \frac{4\en\omega}{\pi}  {\rm{Im}} \chi(\omega) \Theta(\omega,T) \frac{1}{\omega} {\rm{Re}}\int {\rm{d}} V  \left(\bm{E}^{J_x}(\bm{r_v})\right)^{*} \bm{E}^{M_y}(\bm{r_v}). \\
    \label{eq:F}
\end{equation}

Suppressing all position arguments of the relevant fields in a vector notation, 
\begin{align}
{\rm{Re}}\int {\rm{d}} V \left(\bm{E}^{J_x}(\bm{r_v})\right)^{*} \bm{E}^{M_y}(\bm{r_v}) &= {\rm{Re}}\left(\bm{E}^{J_x}\right)^{\dagger}\bm{E}^{M_y} \\ 
& = \frac{1}{\en^2|\chi|^2}  {\rm{Re}}\left(\left(\eincjx  \right)^{\dagger} \mathbb{T}^{\dagger}\mathbb{O}\mathbb{T} \eincmy\right) \\
& = \frac{1}{\en^2|\chi|^2} {\rm Tr} \left( \mathbb{T}^{\dagger}\mathbb{O}\mathbb{T} \; {\rm{Re}}\left(\eincmy \left(\eincjx  \right)^{\dagger}  \right)\right), \\
& = \frac{1}{\en^2|\chi|^2} {\rm Tr} \left( \mathbb{T}^{\dagger}\mathbb{O}\mathbb{T} \; \EE  \right),
    \label{eq:vector_notation}
\end{align}
where $\mathbb{O}$ is a diagonal matrix whose only nonzero diagonal entries that equal 1 correspond to the bottom body (where the thermal sources of interest come from) and the rest of the diagonal entries that equal 0 correspond to the top body. Both $\mathbb{T}$ matrix and ${\bm E}_{\rm inc}$ vectors are defined across both the top and bottom bodies. This $\mathbb{O}$ matrix implies that the outer spatial integral is over the bottom body, instead of both bodies. The matrix $\EE = {\rm{Re}}\left(\eincmy \left(\eincjx  \right)^{\dagger}  \right)$ is a rank-2 matrix that can be decomposed into one positive eigenvalue term and one negative eigenvalue term:
\begin{align}
\EE & = \lambda_1 q_1 q_1^{\dagger} + \lambda_2 q_2 q_2^{\dagger},
\label{eq:EE}
\end{align}
with eigenvalues
\begin{align}
\lambda_{1,2} &= \pm \frac{|\eincjx||\eincmy|}{2}\\
& = \pm \frac{1}{\en} \frac{3.45\times10^{16}}{\left(d\times10^{9}\right)^2} ,
\end{align}
and eigenvectors
\begin{align}
q_{1,2} = \frac{\eincjx}{\sqrt{2}|\eincjx|} \pm \frac{\eincmy}{\sqrt{2}|\eincmy|}.
\end{align}
One can now see that our choice of source amplitudes in \eqref{JxMy_correlation} leads to frequency-independent incident-field eigenvalues.

To bound the expression of \eqref{vector_notation}, we will relax it in a few ways. (Interestingly, intensive numerical optimizations using manifold-optimization techniques~\cite{Absil2009,manopt} directly on \eqref{vector_notation} lead to the same upper limits that we derive below, suggesting that these ``relaxations'' are minimal and do not loosen the analysis given the information, such as sum rules, that we use.) First, the $\EE$ matrix defined by the two renormalized incident fields has one positive and one negative eigenvalue, per \eqref{EE}. Physically, we can interpret the negative sign of the second eigenvalue via the power expression of \eqref{vector_notation} containing $\EE$, as the \emph{difference} in powers absorbed for the two renormalized incident fields. This is of course bounded above by the absorption of only the first incident field, dropping the subtracted term, leaving only the contribution of the single positive eigenvalue of $\EE$. Thus we have:
\begin{align}
\mathbb{T}^{\dagger}\mathbb{O}\mathbb{T} \; {\rm{Re}}\left(\eincmy \left(\eincjx  \right)^{\dagger}  \right) = \mathbb{T}^{\dagger}\mathbb{O}\mathbb{T} \EE \leq  \mathbb{T}^{\dagger}\mathbb{O}\mathbb{T} \; \lambda_1 q_1 q_1^{\dagger}.
\end{align}
Next, we note that $\OO$ indicates absorption only in the lower body; of course this quantity is bounded above by the total absorption in both bodies. Mathematically, the $\OO$ matrix is bounded above by the identity matrix, $\II$. Finally, the absorption in both bodies is less than the net extinction of the two bodies (their far-field scattered powers are positive, and essentially zero in the near-field case, so that this relaxation is negligible), giving:
\begin{align}
{\rm Tr} \left( \mathbb{T}^{\dagger}\mathbb{O}\mathbb{T} \; \lambda_1 q_1 q_1^{\dagger} \right) 
& = \lambda_1 {\rm Tr} \left( q_1^{\dagger} \mathbb{T}^{\dagger}\mathbb{O}\mathbb{T} q_1 \right) \\
& \leqslant \lambda_1 {\rm Tr} \left( q_1^{\dagger} \mathbb{T}^{\dagger}\mathbb{T} q_1 \right),
\end{align}
which is quadratic in $\mathbb{T}$. We can use ``optical theorem'' constraint to bound this quadratic absorption-like quantity with a linear extinction-like quantity. The idea is that absorption must be smaller than extinction: $P_{\rm abs} \leqslant P_{\rm ext}$. Absorption is given in terms of $\mathbb{T}$-matrix by $P_{\rm abs} = \frac{\omega}{2}{\rm Im}\left( {\bm E}^{\dagger}{\bm P} \right) = \frac{\omega}{2\en}\frac{{\rm Im}\chi}{|\chi|^2} {\bm e_{\rm inc}}^{\dagger}\mathbb{T}^{\dagger}\mathbb{T}{\bm e_{\rm inc}}$. Similarly extinction is given by $P_{\rm ext} = \frac{\omega}{2}{\rm Im}\left( {\bm e_{\rm inc}}^{\dagger}{\bm P} \right) = \frac{\omega}{2} {\bm e_{\rm inc}}^{\dagger}{\rm Im}\mathbb{T}{\bm e_{\rm inc}}$. Thus the ``optical theorem'' condition implies that for any $\mathbb{T}$ matrix,
\begin{align}
\frac{{\rm Im}\chi}{\en|\chi|^2}\mathbb{T}^{\dagger}\mathbb{T} \leqslant {\rm Im}\mathbb{T}.
   \label{eq:optical_tm}
\end{align}
For our problem, scattering power is negligible for the entire system, as the near-field heat exchange dominates over the scattering to outside the system. In other words, $P_{\rm abs} \approx P_{\rm ext}$. Thus we can write
\begin{align}
\lambda_1 {\rm Tr} \left( q_1^{\dagger} \mathbb{T}^{\dagger}\mathbb{T} q_1 \right)
& \leqslant \lambda_1 \frac{\en|\chi|^2}{{\rm Im}\chi}{\rm Tr} \left( q_1^{\dagger} {\rm Im}\mathbb{T} q_1 \right),
\end{align}
without introducing much relaxation.

We can now rewrite \eqref{F} as
\begin{align}
F^{\rm{RHT}}(\omega,T) 
& \leqslant \frac{4\en\omega}{\pi}  {\rm{Im}} \chi(\omega) \Theta(\omega,T) \frac{1}{\omega} \frac{1}{\en^2|\chi|^2}\lambda_1 \frac{\en|\chi|^2}{{\rm Im}\chi}{\rm Tr} \left( q_1^{\dagger} {\rm Im}\mathbb{T} q_1 \right) \\
& = \frac{4 \lambda_1}{\pi} \Theta(\omega,T) {\rm Tr} \left( q_1^{\dagger} {\rm Im}\mathbb{T} q_1 \right).
\end{align}
Surprisingly, the various transformations to this point have removed all explicit dependencies on material susceptibility $\chi_{1,2}(\omega)$, with the only implicit dependence embedded in ${\rm Im} \mathbb{T}$. We will now focus on the upper bound for HTC, and the upper bound for RHT can be found by taking similar steps. To switch from the RHT to HTC bound computation, we just need to take the temperature derivative of the last expression to get
\begin{align}
F^{\rm{HTC}}(\omega,T) 
& \leqslant \frac{4 \lambda_1}{\pi} \dtheta {\rm Tr} \left( q_1^{\dagger} {\rm Im}\mathbb{T} q_1 \right),
\label{eq:F_ImT}
\end{align}
where, for simplicity of notation, we can write $x = \frac{\hbar\omega}{\kb T}$ so that
\begin{align}
\dtheta = \kb \frac{x^2 e^x}{\left(e^x-1\right)^2}.
\end{align}

The oscillator representation of the $\mathbb{T}$ matrix, per \eqref{Trep0}, is:
\begin{align}
\mathbb{T} (\omega)&= \sum_{i=1} \frac{\alpha \wi^2}{\wi^2-\omega^2-i\omega\gamma} \mathbb{T}_i = \sum_{i=1} f_i (\omega)\mathbb{T}_i,
\end{align}
where $\mathbb{T}_i$ are the positive-definite oscillator amplitudes and $f_i(\omega) = \frac{\alpha \wi^2}{\wi^2-\omega^2-i\omega\gamma}$ are the oscillator lineshapes. The imaginary part of $\mathbb{T}$ matrix reads
\begin{align}
{\rm Im}\mathbb{T} (\omega)
&= \sum_{i=1} \frac{\en\alpha \wi^2\omega\gamma}{\left(\wi^2-\omega^2\right)^2+\omega^2\gamma^2} \mathbb{T}_i = \sum_{i=1} f_i^{(2)}(\omega) \mathbb{T}_i,
\end{align}
where $f_i^{(2)}(\omega) = \frac{\en\alpha \wi^2\omega\gamma}{\left(\wi^2-\omega^2\right)^2+\omega^2\gamma^2}$. Inserting the oscillator representation into \eqref{F_ImT} we get
\begin{align}
{\rm spectral \; \;HTC} = S_z(\omega,T) &\leqslant \frac{4 \lambda_1}{\pi} \dtheta {\rm Tr} \left( q_1^{\dagger} {\rm Im}\mathbb{T} q_1 \right).
\end{align}
Now we will integrate with respect to frequency, and it will then be obvious that a single $\TT_i$ oscillator at the optimal frequency maximizes the total HTC. For each $\TT_i$ oscillator term, the frequency integral is
\begin{align}
\Omega_i &= \frac{4\lambda_1}{\pi} \int_0^{\infty} {\rm d} \omega \, \dtheta f_i^{(2)}(\omega) \\
&= \frac{4\lambda_1}{\pi} \en\alpha \int_0^{\infty} {\rm d} \omega \,   \frac{\kb x^2 e^x}{\left(e^x-1\right)^2} \omega \frac{\wi^2\gamma}{\left(\wi^2-\omega^2\right)^2+\omega^2\gamma^2}\\
\label{eq:frequency_single}
& = \frac{4\lambda_1}{\pi} \en\alpha \frac{\pi}{2} \frac{\kb^2 T}{\hbar} \frac{x_i^3 e^{x_i}}{\left(e^{x_i}-1\right)^2} ,
\end{align}
where $x_i = \frac{\hbar\omega_i}{\kb T}$. The maximum $\Omega_i$ occurs for $x_i = 2.57$, which corresponds to an optimal oscillator frequency of 
\begin{align}
    \omega_{\rm opt} = \frac{x_0 \kb T}{\hbar},
    \label{eq:omegaopt}
\end{align}
which is exactly the near-field Wien frequency that we found from HTC optimization for planar, unpatterned geometries~\cite{NFRHT_planar}! The maximum $\Omega_i$ at this frequency is
\begin{align}
\Omega_i^{\rm opt} = 2 \lambda_1 \en\alpha \frac{\kb^2 T}{\hbar} \times 1.52 = 0.21 \frac{\kb^2 T}{\hbar d^2}
\label{eq:OmegaOpt}
\end{align}
Apart from the frequency integral, everything else is in the trace, where $\TT_i$ is constrained by $\TT_i >0$ and $\sum_i \TT_i = \II$. The answer to achieving maximum $\int_0^{\infty}  {\rm d} \omega \, S^{\rm{HTC}}(\omega,T)$ is obviously allocating all possible $\TT_i$ to the optimal oscillator frequency. It is easy to find the maximal value of the trace term, as
\begin{align}
{\rm Max}\left( {\rm Tr}  \left( \TT_i q_1 q_1^{\dagger}\right) \right) = 1
\end{align}
when $\TT_i = \II$ or $q_1 q_1^{\dagger}$.
Then the bound will be the product of the frequency integral term and the trace term, both of this optimal oscillator:
\begin{align}
    {\rm HTC} &\leq 0.21 \frac{k_B^2}{\hbar} \frac{T}{d^2} \nonumber \\
              &\leq \beta \frac{T}{d^2},
\end{align}
where $\beta = \SI{3.8e5}{W nm^2/m^2/K^2}$. For $T = 300 \, {\rm K}$ and $d = 10 \, {\rm nm} $, ${\rm HTC} \leq 1.1\times10^{6}  \; {\rm W/m^2/K}$, which is 5X the optimal planar performance. Hence this theoretical framework offers a close prediction to the best known designs, it predicts the optimal resonance frequency where the oscillator-strength should be concentrated, and it explain why previous material-dependent predictions were incorrect.

\bibliographystyle{ieeetr}
\bibliography{tmatrix}